\documentclass[12pt]{iopart}
\usepackage{epsfig}
\usepackage{epsf}
\usepackage{graphicx}
\usepackage{iopams}

\begin{document}

\newcommand{\be}{\begin{equation}}
\newcommand{\ee}{\end{equation}}
\newcommand{\bea}{\begin{eqnarray}}
\newcommand{\eea}{\end{eqnarray}}
\newcommand{\nnn}{\nonumber \\}
\newcommand{\ba}{\begin{array}}
\newcommand{\ea}{\end{array}}
\newcommand{\xv}{{\bf v}}
\newcommand{\xc}{{\bf c}}
\newcommand{\xD}{{\bf D}}
\newcommand{\bx}{{\bf x}}
\newcommand{\hx}{\hat{x}}
\newcommand{\hy}{\hat{y}}
\newcommand{\hz}{\hat{z}}
\newcommand{\htt}{\hat{t}}
\newcommand{\ex}{{\bf e}_x}
\newcommand{\ey}{{\bf e}_y}
\newcommand{\ez}{{\bf e}_z}
\newcommand{\et}{{\bf e}_{\theta}}
\newcommand{\er}{{\bf e}_r}
\newcommand{\xao}{\alpha_0}
\newcommand{\xo}{\mathbf{\omega}}
\newcommand{\xO}{\mathbf{\Omega}}
\newcommand{\xS}{\mathbf{\Sigma}}
\newcommand{\xep}{\langle {\bf p} \rangle}
\newcommand{\xe}{{\bf e}}
\newcommand{\xE}{{\bf E}}
\newcommand{\xI}{{\bf I}}
\newcommand{\bz}{{\bar{z}}}
\newcommand{\xA}{{\bf A}}
\newcommand{\xV}{{\bf V}}
\newcommand{\xk}{{\bf k}}
\newcommand{\xu}{{\bf u}}
\newcommand{\xq}{{\bf q}}
\newcommand{\xx}{{\wedge}}
\newcommand{\xp}{{\bf p}}
\newcommand{\xn}{{\bf n}}
\newcommand{\xj}{{\bf j}}
\newcommand{\xR}{{\bf R}}
\newcommand{\xap}{\langle{\bf p}\rangle}
\newcommand{\xw}{{\bf w}}
\newcommand{\pd}{{\partial}}
\newcommand{\pdt}{{\partial_{t}}}
\newcommand{\pdT}{{\partial_{T}}}
\newcommand{\pdz}{{\partial_{z}}}
\newcommand{\pdx}{{\partial_{x}}}
\newcommand{\pdX}{{\partial_{X}}}
\newcommand{\intab}{\int^0_{-d} dz~~}
\newcommand{\ct}{{\cal T}}
\newcommand{\cF}{{\cal F}}
\newcommand{\tM}{\tilde{M}}
\newcommand{\tN}{\tilde{N}}
\newcommand{\hM}{\hat{M}}
\newcommand{\hN}{\hat{N}}
\newcommand{\OA}{{\cal F}}
\newcommand{\tA}{\tilde{A}}
\newcommand{\tB}{\tilde{B}}
\newcommand{\tO}{\tilde{\Omega}}
\newcommand{\tP}{\tilde{\phi}}
\newcommand{\hA}{\hat{A}}
\newcommand{\hB}{\hat{B}}
\newcommand{\hO}{\hat{\Omega}}
\newcommand{\hP}{\hat{\phi}}
\newcommand{\DO}{D_{\Omega}}
\newcommand{\Sc}{S_c^{-1}}
\newcommand{\nb}{{\bf \nabla}}
\newcommand{\lapl}{{\nabla^2}}
\newcommand{\xJ}{{\bf J}}
\newcommand{\dela}{{\delta_A}}
\newcommand{\del}{{\delta}}
\newcommand{\hf}{\frac{1}{2}}
\newcommand{\degrees}{$^{\circ}$}
\newcommand{\texto}{\rm \large}
\newcommand{\grados}{$^{\circ}C$}
\newcommand{\nl}{\hspace{0.5cm}}
\newcommand{\nota}{ \bf }

\newcommand{\Pe}{\mbox{Pe}}
\newcommand{\RRe}{\mbox{Re}}
\newcommand{\Gr}{\mbox{Gr}}
\newcommand{\Ra}{\mbox{Ra}}

\title[Sheared bioconvection in a tube]{Sheared bioconvection in a horizontal tube}

\author{O. A. Croze, E. E. Ashraf and M. A. Bees}

\address{Department of Mathematics, University of Glasgow, Glasgow G12 8QW, U.K.}
\ead{o.croze@maths.gla.ac.uk}
\begin{abstract}
The recent interest in using microorganisms for biofuels is motivation enough to study bioconvection and cell dispersion in tubes subject to imposed flow.
To optimize light and nutrient uptake, many microorganisms swim in directions biased by environmental cues (e.g. phototaxis in algae and chemotaxis in bacteria).
Such taxes inevitably lead to accumulations of cells, which, as many microorganisms have a density different to the fluid, can induce hydrodynamic instabilites.
The large-scale fluid flow and spectacular patterns that arise are termed bioconvection.
However, the extent to which bioconvection is affected or suppressed by an imposed fluid flow, and how bioconvection influences the mean flow profile and cell transport are open questions.
This experimental study is the first to address these issues by quantifying the patterns due to suspensions of the gravitactic and gyrotactic green biflagellate alga \emph{Chlamydomonas} in horizontal tubes subject to an imposed flow.
With no flow, the dependence of the dominant pattern wavelength at pattern onset on cell concentration is established for three different tube diameters.
For small imposed flows, the vertical plumes of cells are observed merely to bow in the direction of flow.
For sufficiently high flow rates, the plumes progressively fragment into piecewise linear diagonal plumes, unexpectedly inclined at constant angles and translating at fixed speeds.
The pattern wavelength generally grows with flow rate, with transitions at critical rates that depend on concentration.
Even at high imposed flow rates, bioconvection is not wholly suppressed and perturbs the flow field.
\end{abstract}

\maketitle

\section{Introduction}

Most populations of motile microorganisms bias their motion in response to environmental cues to seek better conditions
(e.g. bacterial chemotaxis, Berg 2004).  To find water surfaces, phototrophic green algae, such as {\it Chlamydomonas} and {\it Dunaliella}, are gravitactic (they swim upwards on average).  Typically the upward bias is due to a combination of the cell's centre-of-mass being displaced from its centre-of-buoyancy and a torque due to sedimentation of the asymmetric body-flagella ensemble.
Additionally, such cells are often gyrotactic (on average they swim towards downwelling regions due to a balance of viscous and gravitational torques; Kessler 1985) and phototactic (biased motion relative to the illumination; Foster \& Smyth 1980).
The taxes invariably lead to accumulations of the negatively buoyant cells, inducing hydrodynamic instabilities that result in spatially localized structures and patterns, termed bioconvection (see Pedley \& Kessler 1992; Hill \& Pedley 2005).
In particular, overturning instabilities occur when cells accumulate at the upper surface, forming a dense, unstable layer (analogous to Rayleigh-Bernard convection; Chandrasekar 1961), yielding patterns of spots, stripes or more complicated structures.
Away from the upper boundary, gyrotactic instabilities lead to long thin plume structures, where cells swim towards downwelling regions and their added mass amplifies the downwelling.

Recently, Durham {\it et al.}~(2009) showed that transient gyrotactic trapping of motile phytoplankton in differentially sheared fluids can generate thin, cell-rich layers in the ocean, with important consequences for the ecology of toxic algae. Other
investigations (e.g. Bearon \& Gr\"{u}nbaum 2006; Gr\"{u}nbaum 2009) demonstrate the ecological importance of spatially localized structures (patchiness, due to cell swimming).
In biotechnology applications, the properties of swimming cell suspensions have been minimally exploited or outright ignored.
Recently, there has been renewed interest in the production of biofuel from microorganisms
(especially biodiesel or hydrogen; Melis \& Happe 2001; Chisti 2007). For these alternative fuels to become commercially competitive, existing bioreactor designs need to be refined. In current closed bioreactors, algae are cultured in laminar or turbulent flows through arrangements of horizontal, slanted and/or vertical tubes, bubbled for gas exchange, and concentrated by flocculation, filtration or centrifugation, as if they were chemicals or inert colloids (Grima {\it et al.} 2001, 2003;  Garc\'{i}a-Gonz\'{a}lez \etal 2005; Chisti 2007). However, candidate species for biofuel production that swim (e.g.~{\it Dunaliella}) do not behave like chemicals or inert colloids: their peculiar collective motions and consequent transport properties need to be taken into account, and indeed exploited.
Whether the flow in tubes is laminar or turbulent, gyrotactic swimming cells organize in patterns that alter the flow.

Bees \& Croze (2010) analysed the dispersion of swimming cells in a flow in a vertical tube in the laminar regime. They predicted the effective drift and diffusion of gyrotactic algae in plumes; depending on the algal cell properties, algal dispersion can be quantitatively very different from that of classical Taylor-Aris dispersion for inert chemicals. Bees \& Croze (2010) also obtained steady state plume solutions, which show that the flow in the presence of cells deviates from the standard Poiseuille profile.

It is well known that the transition to turbulence in a tube is strongly sensitive to the initial laminar state (Willis \etal 2008). It is interesting to note that the perturbation by the presence of swimming cells should inevitably change the onset of the transition. Cells in turbulent flows in pipes have not been analysed, but Lewis (2003) recently showed that gyrotactic algae in a homogenous and isotropic turbulent flow field retain their bias. Turbulence only changes the effective value of the diffusivity of cell orientation.

\begin{figure}
\begin{center}
(a)\,\includegraphics[width=0.6\linewidth]{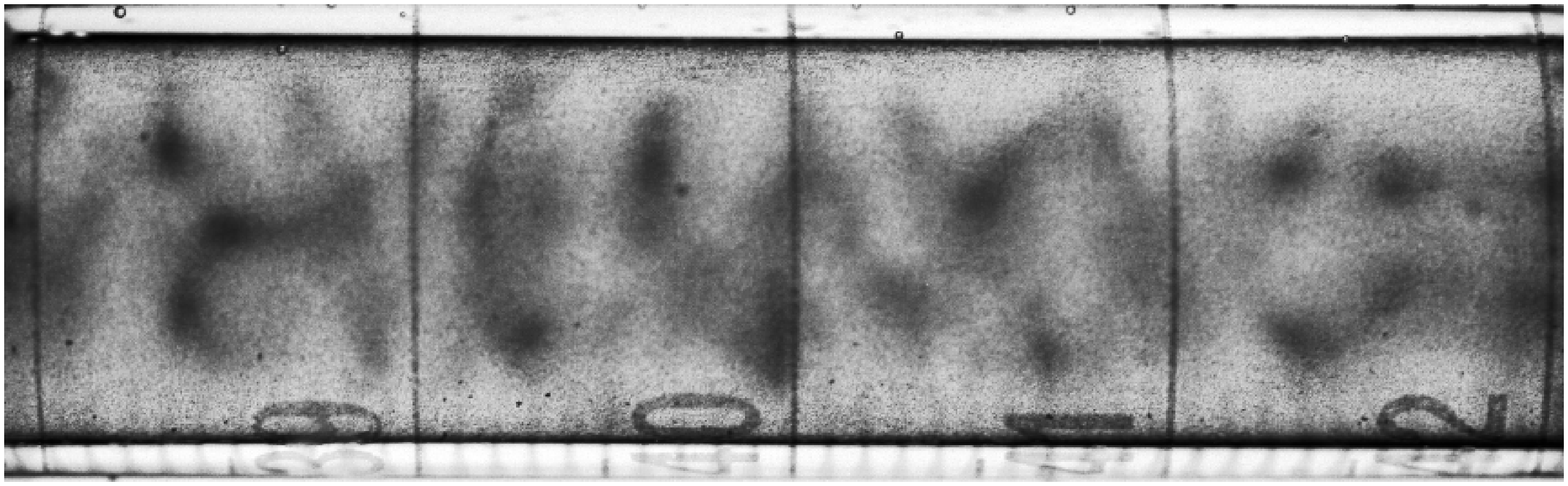}\\
(b)\,\includegraphics[width=0.6\linewidth]{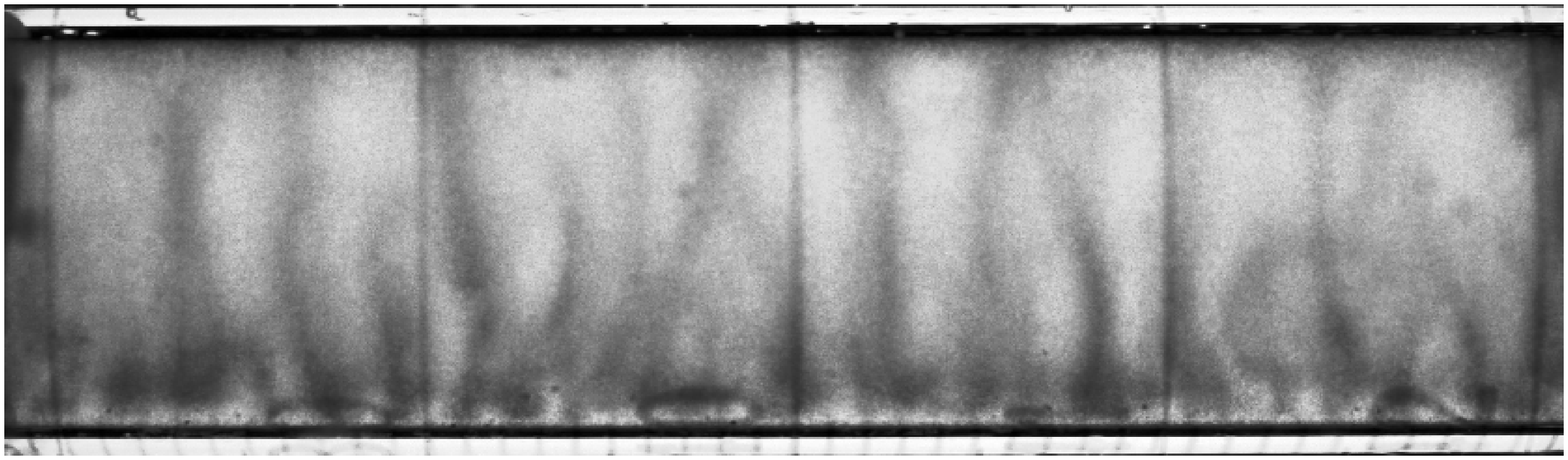}
\caption{Bioconvection pattern of \emph{C. augustae} in a tube (tube A; concentration, $c\approx1.4\times10^6$ cells/ml) $8$ minutes after the onset of instability, as imaged: (a) from the top, focusing close to the top of the tube; (b) from the side, focusing on the center of the tube. The observed patterns are similar to those originally photographed by Wager (1911) for \emph{E. viridis}. The images are $4.2$ by $1.3$ cm.}\label{topside}
\end{center}
\end{figure}

In this paper, we present an experimental exploration of bioconvection in horizontal tubes under conditions of no imposed flow and in the presence of weak shear. We aim to investigate the transport of algae in a minimally perturbed system with bioreactor geometry, in the same spirit as Bees \& Croze (2010), and with a view to understanding the coupled cell and fluid dynamics in bioreactors. This is the first study of bioconvection under shear, and the first {\it quantitative} study of bioconvection in horizontal tubes. The only other experimental study of bioconvection in horizontal tubes is the pioneering work by Wager (1911). Wager placed suspensions of the alga {\it Euglena viridis} in $25$ cm long tubes of diameter $0.7$ cm, and described and imaged the resulting `aggregations' of cells  (see figure \ref{topside}): an instability was visible after $30$ s and the patterns evolved to a stable state over tens of minutes; cell concentrations were not established, but patterns were sharper and formed quicker for more concentrated samples. Wager repeated the experiments with {\it Chalmydomonas} cells ($20$ cm long tubes of diameter $0.8$ cm), obtaining similar structures (but requiring larger concentrations).

There are few quantitative experimental studies of bioconvection and these have been limited to shallow layers with a free upper surface. Bees \& Hill (1997) analysed bioconvection patterns of {\it Chlamydomonas nivalis} (in this paper denominated {\it Chlamydomonas augustae}, see methods) in $5$ cm diameter Petri dishes. From Fast Fourier Transforms (FFT) of processed pattern images they obtained the concentration, depth and time dependence of the initial and long-term pattern wavelengths. Both wavelengths were found to fall off with concentration. The initial wavelength also increased with depth (see Czir\'{o}k \etal 2000),
while the final wavelength did not strongly depend on it (due to gyrotactic focusing).
Using a similar approach, the wavelengths of the bioconvection patterns due to the aerotactic bacterium {\it B. subtilis} have been quantified
(J\'{a}nosi \etal 1998; Czir\'{o}k \etal 2000). The initial pattern wavelength depended on cell concentration but not depth. The differences between species were ascribed to the particular taxes in operation: aerotactic bacteria are in thin layers where oxygen gradients exist, whilst gravity affects gyrotactic cells throughout the layer.

Theoretical studies of bioconvection are more numerous. For green algae, continuum models of the coupled fluid flow and cell concentration have evolved significantly from deterministic models of pure gravitaxis (Levandowsky \etal 1975, Childress \etal 1975) to stochastic models of gyrotactic cells (Pedley \& Kessler 1990). Predictions for the wavelengths at the onset of bioconvection from linear stability analyses give values of the same order of magnitude as experiments: slightly larger for uniform suspensions (Pedley \& Kessler 1990) and smaller for layers of finite depth (Bees \& Hill 1998). The discrepancies between theory and experiment have been attributed to the uncertainty with which key experimental parameters are known, nonlinear effects and to the non-attainment of equilibrium before instability arises (\textit{idem}). Nonlinear effects have been explored by Bees \& Hill (1999) who calculated static and travelling plume solutions for bioconvection in deep layers.  We refer the reader to the reviews by Pedley \& Kessler (1992) and Hill \& Pedley (2005) for further details.

In the next section, we describe the cultures, experimental set-up and data acquisition methods used in this work. In section \ref{results} we present results quantifying bioconvection patterns first in the absence and then in the presence of imposed flow. Finally, in section \ref{disc} we discuss these results and draw conclusions from our observations.

\section{Materials and methods \label{matmed}}

\subsection*{Cell culture and concentration}

Batch cultures of {\it C. augustae} (CCAP 11/51B; Culture Collection of Algae and Protozoa) were grown statically in conical flasks to exponential phase on 3N-BBM (nitrogen-enriched Bold's basal medium; Schl\"{o}sser 1997) at a stable ambient temperature of $T=24\pm1^{\circ}$C.
The up-swimming cells accumulate on top of sterile cotton wool placed in the neck of a full flask and can be harvested by gentle suction with a Pasteur pipette after 2 or 3 days, and diluted to provide a range of concentrations.
Concentration was measured with a spectrophotometer (WPA CO7500) from readings of $A_{590}$, the absorbance at 590 nm, calibrated with a haemocytometer. Note that the green alga known as {\it C. nivalis} (CCAP 11/51B, see Kessler \etal (1992)), the model organism for bioconvection studies, was recently shown to have been mistakenly identified (Pr\"{o}schold \etal 2001). It has been renamed {\it C. augustae}, and we use this denomination herein.

\subsection*{Set-up, mixing, illumination and flow}

\begin{figure}
\begin{center}
\includegraphics[width=0.7\linewidth]{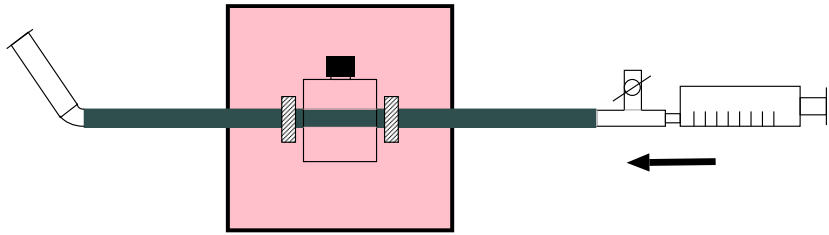}\vspace*{-0.2cm}
\caption{Schematic side-view of the experimental setup. The center of the tube is encased with a glycerol-filled flat-faced perspex bottle to minimize optical distortion, clamped where indicated by the shaded boxes and illuminated by a red LED light source fixed behind the tube. A camera (not shown) mounted in front of the tube captures images from the side. Suspensions are loaded via a plastic tube affixed to the left of the tube. Flow (in the direction of the arrow) is imposed using a syringe pump on the right of the tube, where we also see a tap to facilitate air removal when loading.}\label{expdiag}
\end{center}
\end{figure}

Three tubes of different diameters (pyrex glass or plastic; table \ref{tubechar}) were employed (see figure \ref{expdiag}). The middle of each tube was encased with a flat-faced perspex bottle filled with water or glycerol to minimize optical distortion. Results using glycerol, where distortion is small but non-negligible at the tube edges, are systematically corrected (see below). Tubes were clamped about the central casing, placed on an isolated work bench to minimize vibrations, and levelled before introducing the suspension.
\begin{table}
\caption{Table of diameters and lengths of different cylindrical tubes used in the experiments. Tubes are made of pyrex glass except tube C which is plastic.\label{tubechar}}
\begin{center}
   \begin{tabular}{c  c  c }
   \hline
   tube\,\,\,\,\,\,& inner/outer diameter (cm)\,\,\,\,\,\,  & length (cm)
   \\[1ex]
   \hline
    A  &  1.10 / 1.35 & 30.0\\
    B  &  0.79 / 1.03  & 28.5     \\
    C  &  0.59 / 0.81 & 33.0    
   \\[1ex]
   \hline
   \end{tabular}
\end{center}
\end{table}

Tubing was attached to the ends of the cylindrical tubes to aid mixing and to drive flows (figure \ref{expdiag}). Suspensions were mixed in a plastic bottle using a slow magnetic stirrer before being poured into the tube and imaged. The pouring (and any inadvertent entrainment of bubbles) generated secondary flows, and appeared to aid mixing.  Other mixing protocols gave less satisfactory results. The proportion of swimming cells was reduced by repeated mixing due to deflagellation and adhesion of cells to the boundaries.  To prevent the latter, we tried soaking the tubes in bovine serum albumen (BSA), which limits adhesion in microchannels (Weibel \etal 2005), but patterns were slightly suppressed. The decrease of the concentration by surface fouling was not significant on the experimental timescale ($10$ minutes). It is conceivable that in the longer lasting experiments with flow, the flow itself helped prevent adhesion.

Tubes were illuminated with diffuse red light of wavelength $660$ nm (BL1960 LED light bank; Advanced Illumination, Rochester, USA; at less than 100 lux), to which green algae do not respond phototactically (Nultsch \etal 1971; Foster \& Smyth 1980).

Flow in the tubes was generated using a Graseby 3500 syringe pump (Graseby Medical Ltd., Watford, UK) attached at one end of the experimental setup (figure \ref{expdiag}) to induce a pressure gradient that results in a weak Poiseuille flow if cells are absent. Experiments were performed with flow rates in the range $0-35$ ml/h, as described in detail the next section, at a stable ambient temperature of $T=24\pm1^{\circ}$C.

\subsection*{Data acquisition, processing and analysis}

Timelapse sequences of projected bioconvection patterns were captured from the side and from above using either a computer controlled CL-1014 B/W CCD camera (Camtek) or EOS 350D SLR digital camera (Canon) for higher resolution results. In the absence of flow, up to 201 images were recorded at intervals of 2 to 5 s, whilst for the flow experiments 49 images were recorded a second apart. Images were acquired ($768\times576$ pixels for CL-1014; $3456\times2304$ pixels for EOS 350D) and processed using IDL (RSI, Boulder, USA). The first image of a sequence, if featureless, was subtracted to remove static noise; when features appeared in the first image (flow results), the stack median was subtracted instead.  Each cropped image, $h(x,y)$, was windowed before applying the FFT to give $H(k_x, k_y)$ (see Bees \& Hill 1997), yielding the biased spectrum
\be
I(k)=\sum_{d=k, |\psi| \leq \pi/4} |H(k_x, k_y)|^2,\label{eq:I}
\ee
where $d(k_x, k_y)=\sqrt{k_x^2+k_y^2}$ and $\psi(k_x, k_y)=\arctan(k_y/k_x)$ is the angle to the horizontal. In (\ref{eq:I}), Fourier modes $(k_x,k_y)$ with $|\psi| > \pi/4$ were filtered to eliminate contributions from the boundaries of the tube and cross-sectional structure. The double logarithmic fit proposed by Czir\'{o}k \etal (2000) was employed: $\ln[I(k)]=\alpha|\ln(k)-\ln(k_{0})|-\beta \ln(k)+c$, where $\alpha$, $\beta$, $c$ and $k_0$ (the dominant wavenumber) are fitting parameters.  Wavelengths are given by $\lambda=I_W/k_0$, where $I_W$ is the image width. The initial wavelength, $\lambda_i$, was determined from the fitted Fourier spectra together with direct observation of image sequences and contour plots (figure \ref{seqcontplot}i) as a guide when bimodal distributions (occasionally) occurred. The time for the establishment of the initial instability was measured as the time between pouring in a mixed suspension and measurement of the initial wavelength. The final pattern wavelength, $\lambda_f$, was calculated as an average over the last $50$ images of each sequence; the standard deviation of this sample was taken as an estimate of uncertainty in the value of $\lambda_f$.  When more than one value was obtained for a given concentration, the weighted average and standard error in the mean of the results were reported.

With imposed flow, images were acquired as follows. The cell suspension was mixed and slowly poured in as before. Then, after waiting $3-5$ minutes for the pattern to form, $49$ images were recorded. The flow rate was then increased to $2$ ml/h, and after waiting $2$ minutes, another $49$ frames were recorded. Proceeding similarly, the flow rate was increased in steps of $2$ ml/h up to $10$ ml/h, and then in steps of $5$ ml/h up to a maximum of $35$ ml/h.  (For some experiments only $5$ ml/h steps were used). The dominant wavelength was obtained by averaging over the last $20$ frames of each sequence (the standard deviation was taken as an estimate of the variation). To quantify pattern distortion and plume dynamics, the mean angle the plumes make to the vertical towards the top and bottom of the tube, $\langle \theta \rangle$, and the mean plume drift speed, $\langle V \rangle$, were measured. Plume angle measurements were obtained using a straight line selection tool in ImageJ, NIH, for the approximately straight sections of the plumes. Straight plume fragments were measured directly in the upper or lower regions of the tube or, when curved plumes spanned the whole tube, measurements were taken at the top and bottom boundaries.
Angles were averaged across plumes from three clear images in each sequence (the standard deviation was used as a measure of uncertainty).
For the plume drift speeds, each plume was tracked in a sequence and its displacement, $L_p$, recorded over a total time $T_p$, giving the mean plume speed, $V=L_p/T_p$ (the portion of the plume that does not change shape significantly was tracked; the plumes were not observed to accelerate). The mean of the speeds over different plumes provides $\langle V \rangle$.

For results with the glycerol filled jacket, the image distortion due to refraction at the water-glycerol interface was corrected: if $z'$ is the measured vertical position of a pixel, $z$ is the actual vertical position ($z=0$ at the tube center) and $\nu=1.10$ is the ratio of the index of refraction of pyrex/glycerol to that of water, then approximately $|z|\approx \nu |z'|$. Distortion only operates vertically, and marginally affects the measurement of angles as described above (corrected via $\tan(\langle\theta\rangle) \approx \tan(\langle\theta'\rangle)/\nu$; less than a 3$^{\circ}$ change).

\section{Results \label{results}}

Here we report the results of our observations and quantitative analysis of bioconvection in horizontal tubes, first in the absence of flow and then for sheared bioconvection.

\subsection{No flow}

\begin{figure}
\begin{center}
$\begin{array}{c@{\hspace{.1in}}c}
(a)\,\includegraphics[width=0.3\linewidth]{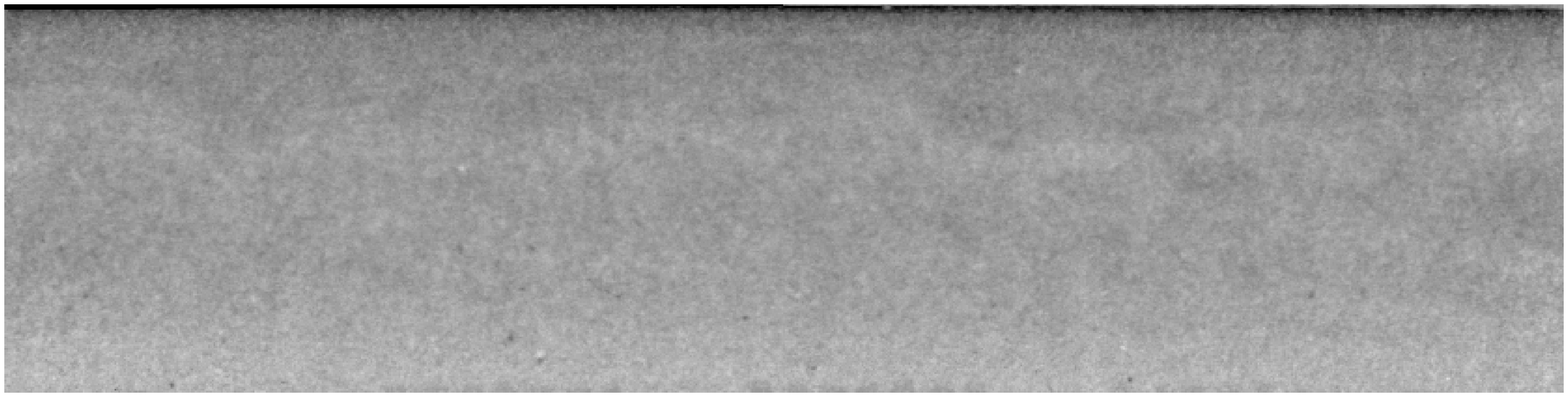}  &  
(e)\,\includegraphics[width=0.3\linewidth]{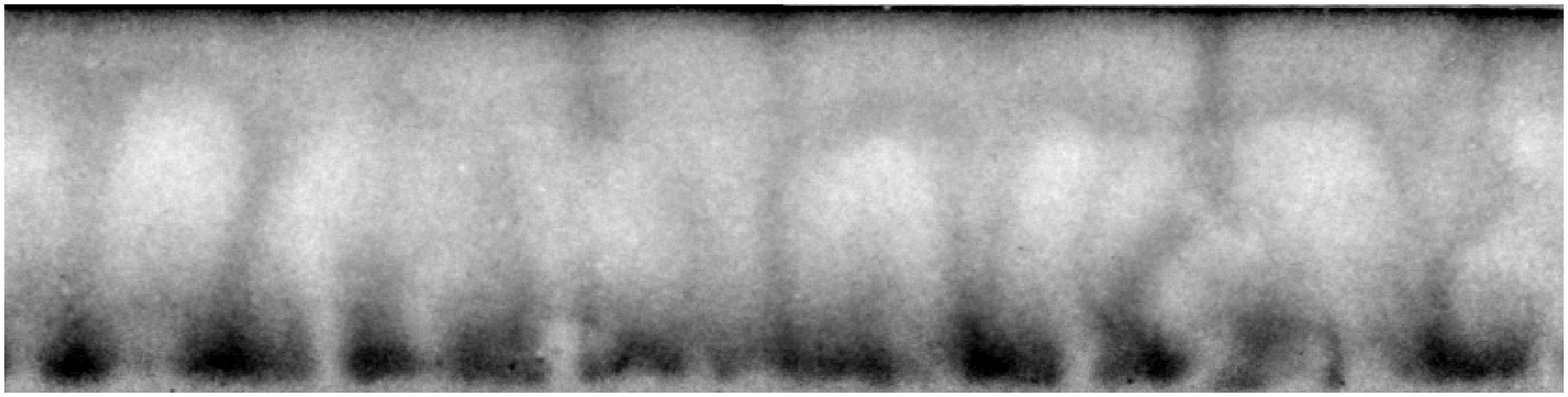}    \\
(b)\,\includegraphics[width=0.3\linewidth]{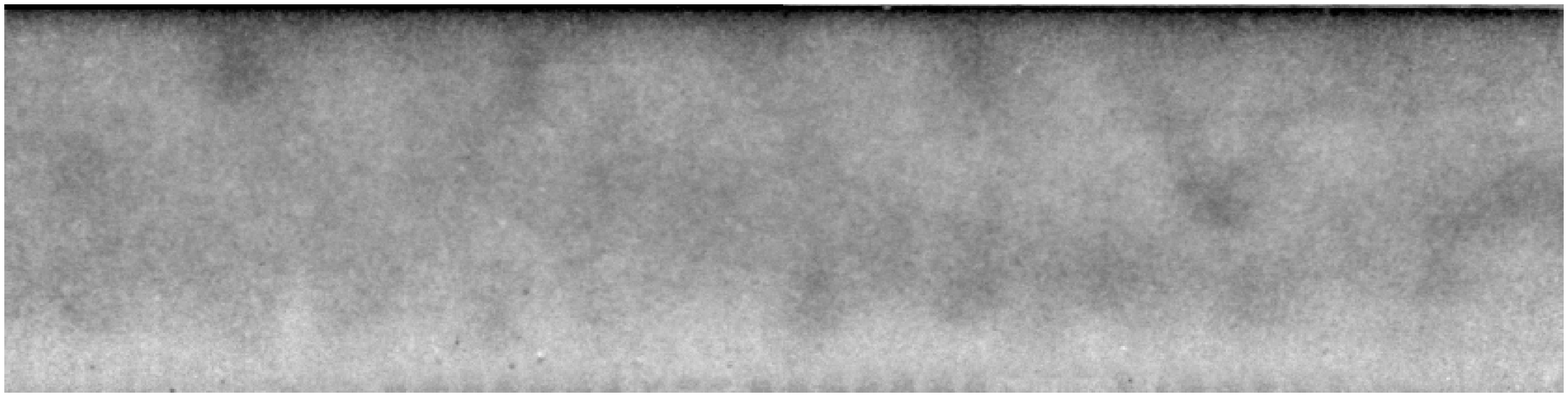}  &  
(f)\,\includegraphics[width=0.3\linewidth]{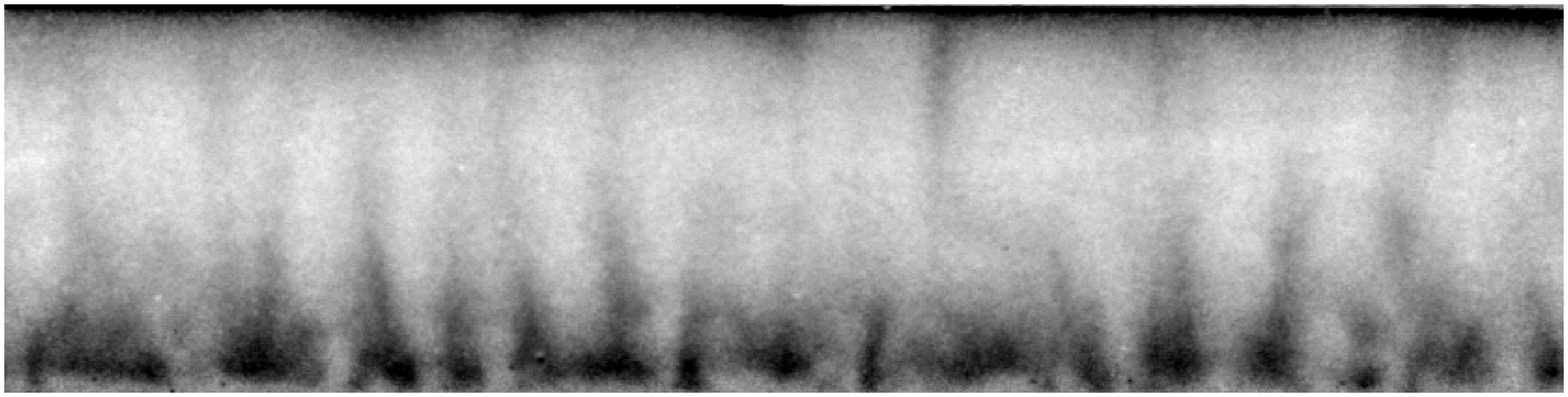}    \\
(c)\,\includegraphics[width=0.3\linewidth]{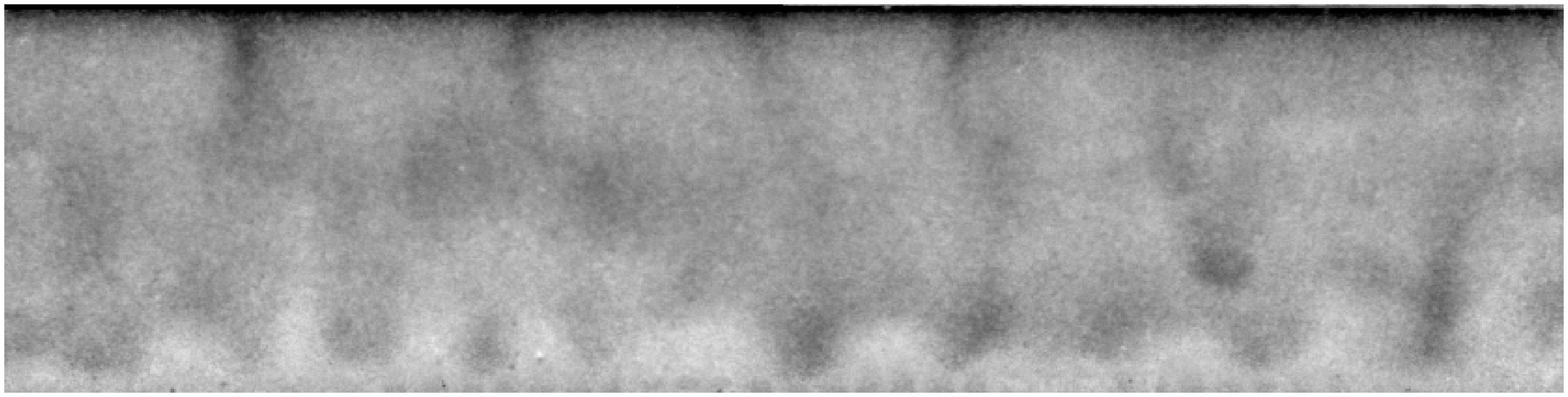}  &  
(g)\,\includegraphics[width=0.3\linewidth]{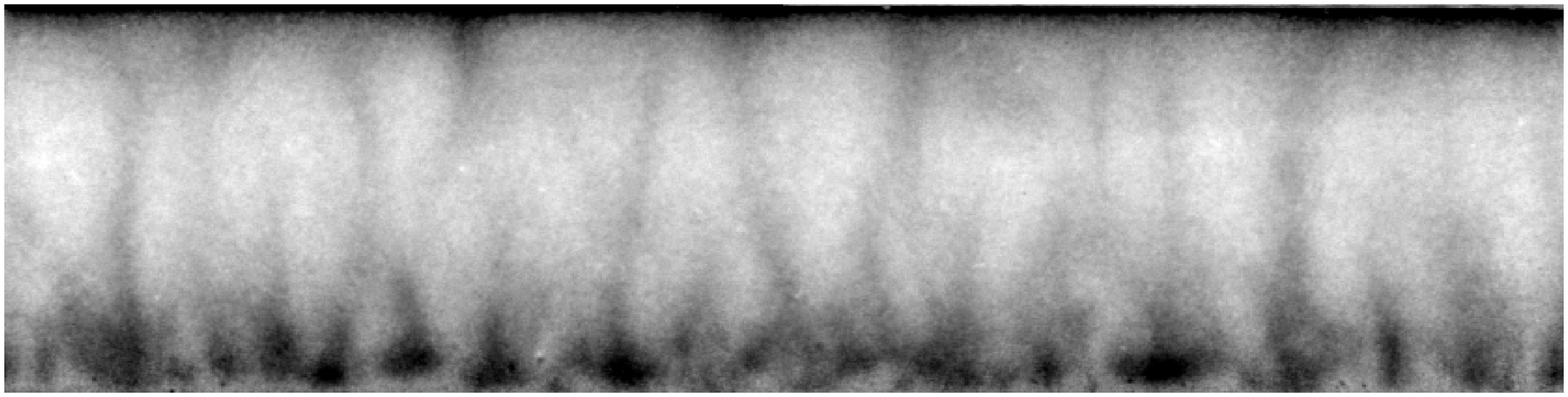}    \\
(d)\,\includegraphics[width=0.3\linewidth]{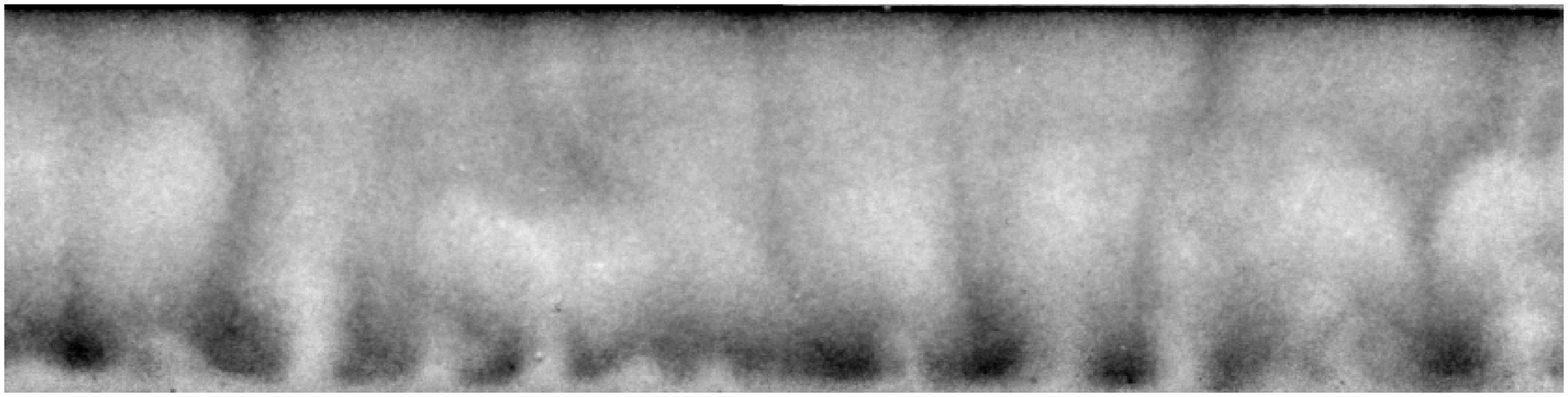}  &  
(h)\,\includegraphics[width=0.3\linewidth]{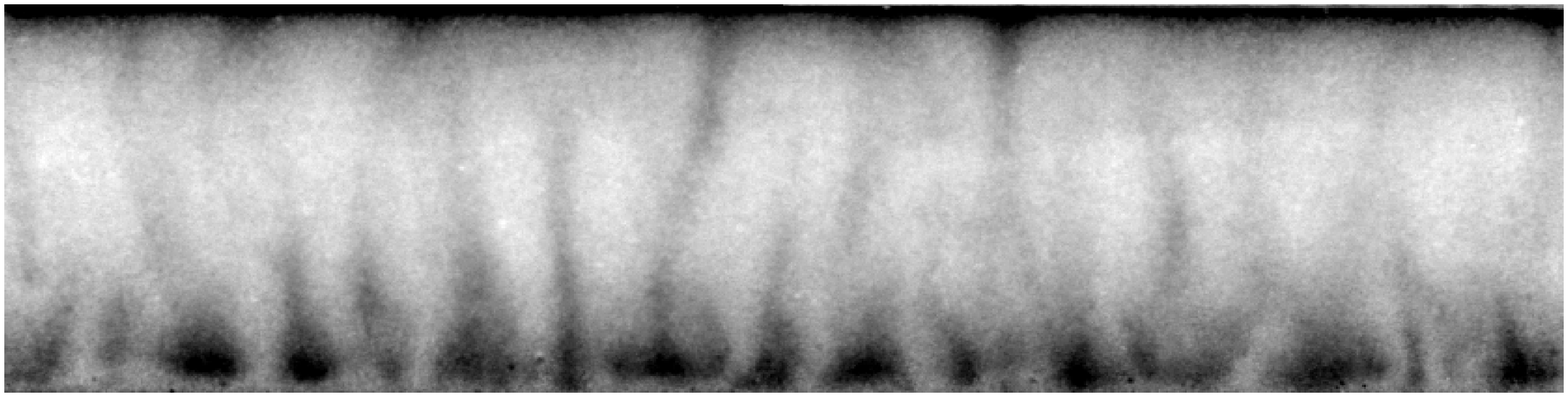}    \\
\end{array}$
(i)\,\includegraphics[width=0.8\linewidth]{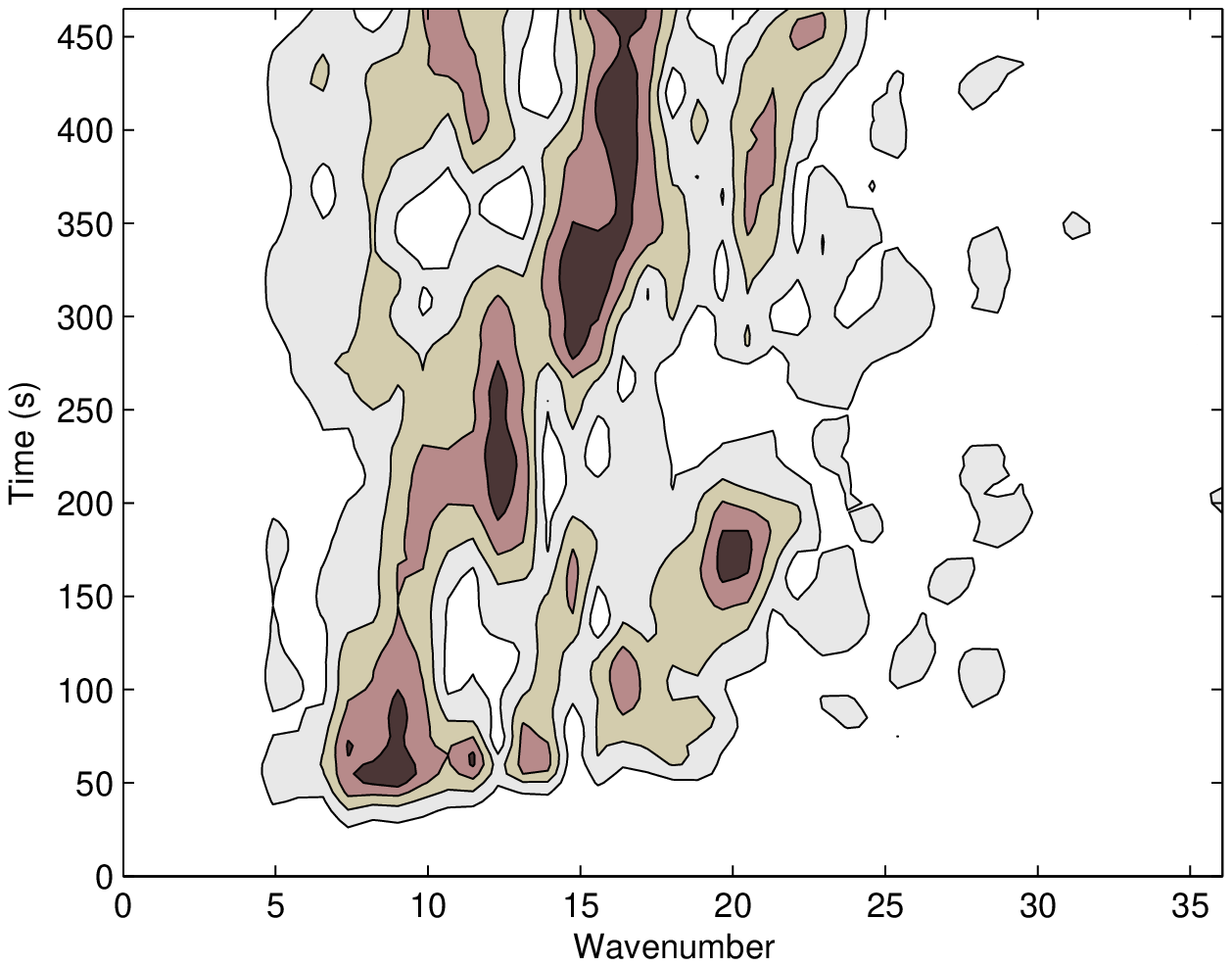}
\caption{Image sequence (a-h) and corresponding time-wavenumber contour plot (i) of the Fourier spectrum density of a representative bioconvection pattern (tube A, $c=1.4\times10^6$ cells/ml). In the sequence the processed images are shown at times 20, 35, 45, 60, 75, 160, 260, 435 s (a-h). The sequence and contour chart the evolution of the pattern from the onset to the final steady state.}\label{seqcontplot}
\end{center}
\end{figure}

In the absence of flow, the pattern development was recorded after a cell suspension was mixed and poured into each tube. Experiments were carried out with tubes A, B and tube C (inner diameters $1.10,0.79$ and $0.59$ cm respectively), varying the cell concentration, $c$, in the range $c=0.5\times10^6-4\times10^6$ cells/ml. The sequence in figure \ref{seqcontplot}(a-h) shows the evolution of a typical bioconvection pattern in a tube as viewed from the side (tube A, $c=1.84\times10^6$ cells/ml). We observe that a well mixed suspension (homogeneous in the first frames) quickly becomes unstable, breaking up into a beautiful striated alternation of thin dark plumes (high cell concentration) and white bands (low cell concentration).
Similarly to bioconvection patterns in other geometries, the pattern is seen to sharpen in time, with plumes becoming spaced closer together. As one would expect for cylindrical tubes of uniform diameter, statistically similar patterns are observed along the length of the tube.

The above observations are usefully quantified by analysis of the dominant Fourier modes extracted from the images (see methods). Figure \ref{seqcontplot}i shows a contour plot of the power spectrum of the image sequence in figure \ref{seqcontplot}(a-h), demonstrating the evolution of pattern wavenumbers with time. The Fourier spectrum quantifies our observations: an initially homogenous suspension quickly suffers an initial instability, with wave number $k_i$ (the first `peak' in figure \ref{seqcontplot}i); the pattern then becomes unstable to other modes, evolving to a structure with an average final wave number $k_f$. The complex dynamics with which modes evolve from the initial instability is also clear in figure \ref{seqcontplot}i.
\begin{figure}
\begin{center}
(a)\includegraphics[width=0.45\linewidth]{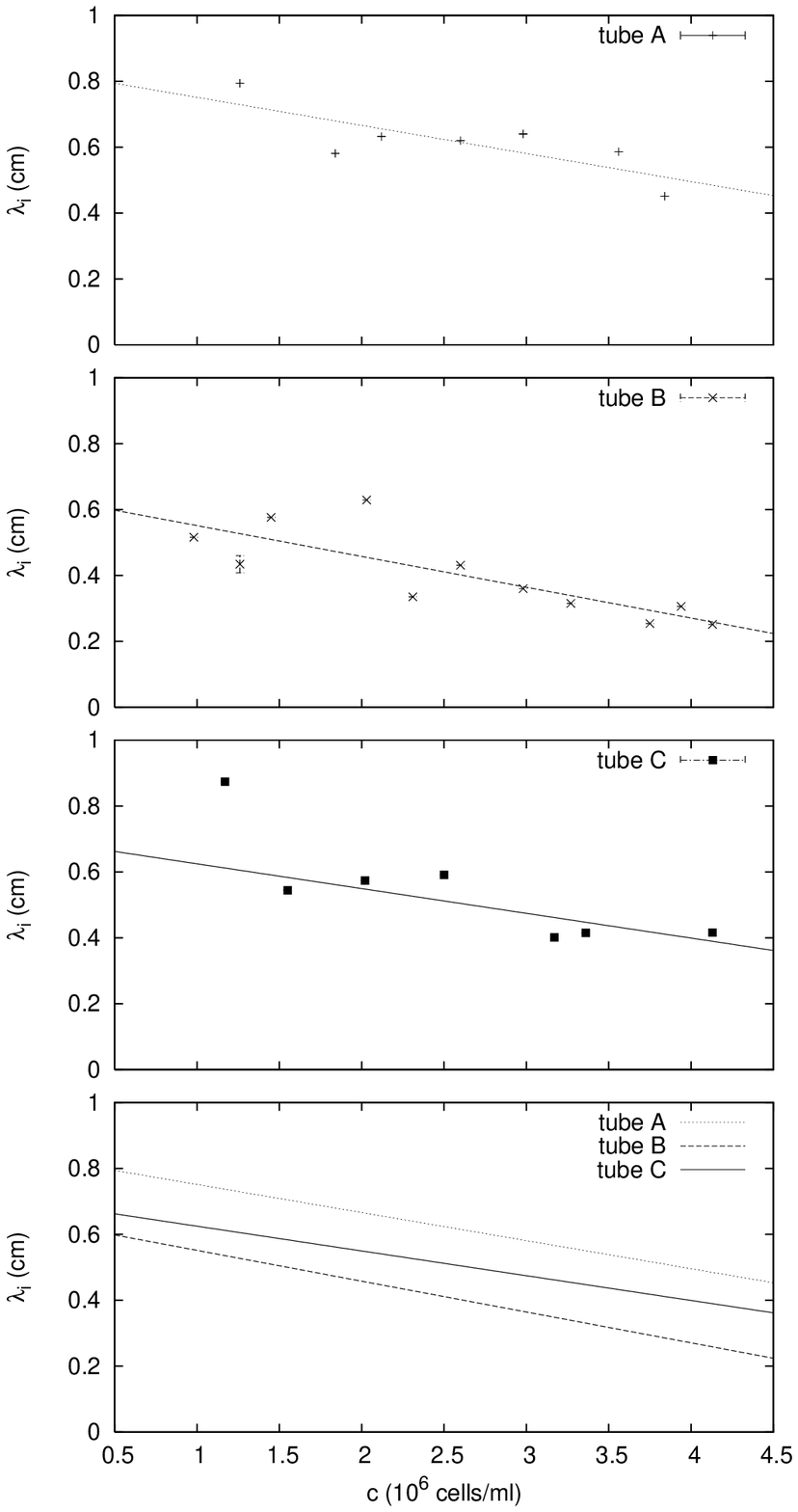}
(b)\includegraphics[width=0.45\linewidth]{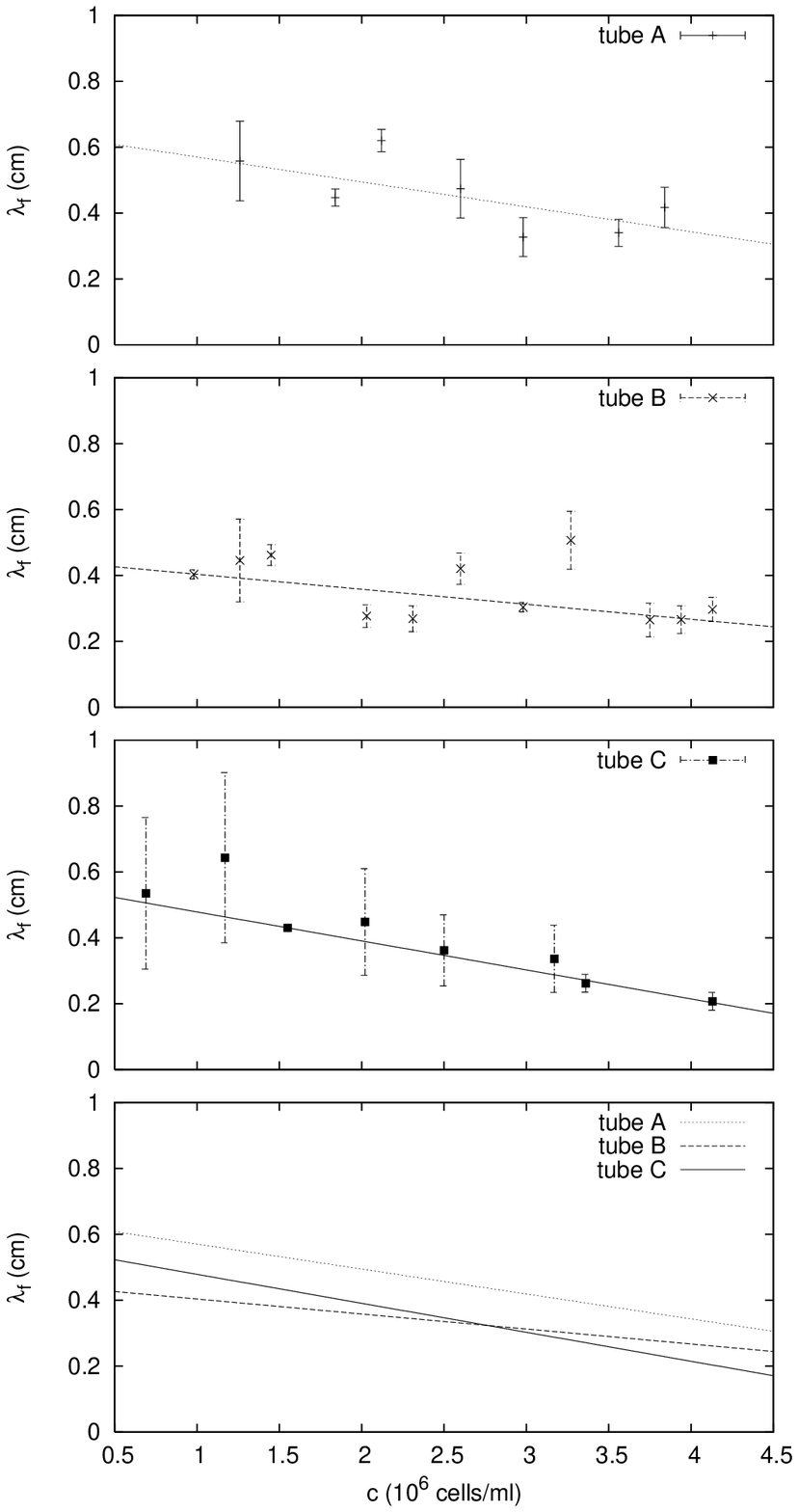}
\caption{Initial (a) and final (b) dominant wavelengths, $\lambda_i$ and $\lambda_f$ respectively, as a function of concentration for tubes A, B and tube C (top three panels, as indicated). Straight line fits to the data are shown (for tube C in (a) the fit was for $c\geq1.5\times10^6$ cells/ml, see text). The last panel shows the fits without the data to allow a comparison of magnitudes. The error bars shown are standard deviations over a sample of $50$ fits in a sequence (b) or the standard error in the mean over repeated experiments (a, b).}\label{infinC}
\end{center}
\end{figure}
\begin{figure}
\begin{center}
\includegraphics[width=0.45\linewidth]{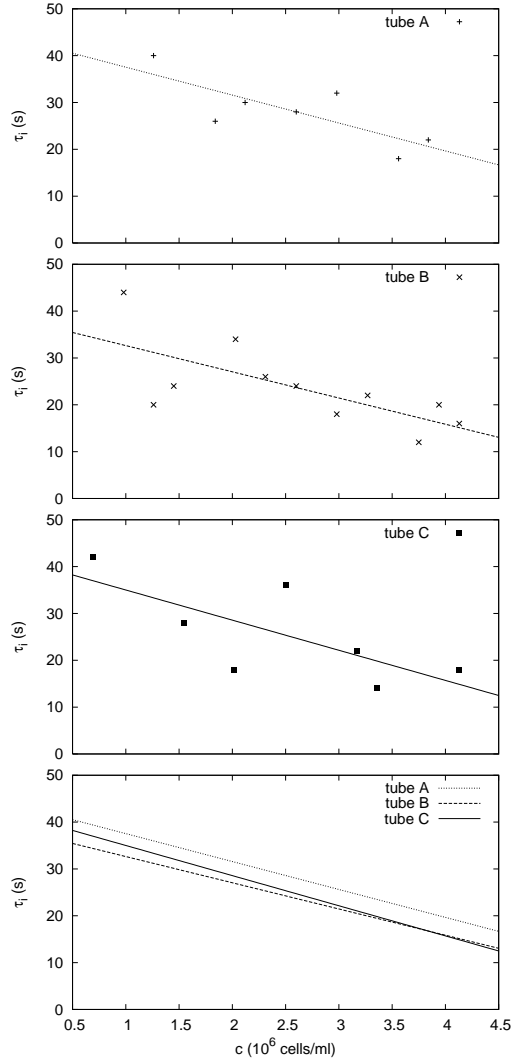}
\caption{Variation of time to establish the initial instability for different cell concentrations for tubes A, B and C (top three panels, as indicated), and  comparison of linear fits to this variation (bottom panel).}\label{pattgrowthtime}
\end{center}
\end{figure}
From figure \ref{infinC}, it is clear that the initial and final dominant pattern wavelengths, $\lambda_i$ and $\lambda_f$ respectively, are consistent with a linear drop with concentration. A possible exception is the narrow tube C, where the large rise in $\lambda_i$ at very low concentrations might be evidence for stronger dependence. Thus, excluding points below $c=1.5\times10^6$ cells/ml in tube C for  $\lambda_i$, the data can be described by the linear form $\lambda(c)=\lambda_0+ \sigma c$, where $\lambda_0$ (intercept) and $\sigma$ (gradient) are the fit parameters. Performing a least squares fit for the initial wavelength: ($\sigma_i^{\rm A}=-0.085\pm0.030$ cm ml/cells,  $\lambda^{\rm A}_{0,i}=0.84\pm0.08$ cm) for tube A;
($\sigma_i^{\rm B}=-0.093\pm0.023$ cm ml/cells,  $\lambda^{\rm B}_{0,i}=0.64\pm0.06$ cm) for tube B; and ($\sigma_i^{\rm C}=-0.075\pm0.028$ cm ml/cells,  $\lambda^{\rm C}_{0,i}=0.70\pm0.08$ cm) for tube C. For the final wavelength we find similar values, ($\sigma_f^{\rm A}=-0.076\pm0.049$ cm ml/cells,  $\lambda^{\rm A}_{f,0}=0.65\pm0.13$ cm) for tube A; ($\sigma_f^{\rm B}=-0.046\pm0.013$ cm ml/cells,  $\lambda^{\rm B}_{f,0}=0.45\pm0.03$ cm) for tube B; and ($\sigma_f^{\rm C}=-0.088\pm0.004$ cm ml/cells,  $\lambda^{\rm C}_{f,0}=0.57\pm0.01$ cm) for tube C. Within the uncertainties of our measurements, it is hard to ascertain any dependence of wavelength on tube diameter, but it would be reasonable to infer that both the initial and final wavelengths are in general larger for tube A than for tube B or C.

Aside from the initial wavelength, another useful measure of the onset of bioconvection is the time $\tau_i$ taken to establish the initial instability (figure \ref{pattgrowthtime}). Like $\lambda_i$, $\tau_i$ falls linearly with concentration, but without any obvious sudden rise at low $c$ for tube C. Using a least squares fit with a function of the form $\tau_i(c)=\tau_{0,i}+ \gamma_i c$, (with $\tau_{0,i}$ and $\gamma_i$ fit parameters, as above), we find: ($\gamma_i^{\rm A}=-6.0\pm2.1$ s ml/cells, $\tau_{0,i}^{\rm A}=43\pm6$ s) for tube A;  ($\gamma_i^{\rm B}=-5.6\pm1.9$ s ml/cells, $\tau_{0,i}^{\rm B}=38\pm5$ s) for tube B; and ($\gamma_i^{\rm C}=-6.4\pm2.7 $ s ml/cells, $\tau_{0,i}^{\rm C}=41\pm7$ s) for tube C.  The dependence on tube radius again is not strong.

\subsection{Effect of flow}

\begin{figure}
\begin{center}
\includegraphics[width=0.8\linewidth]{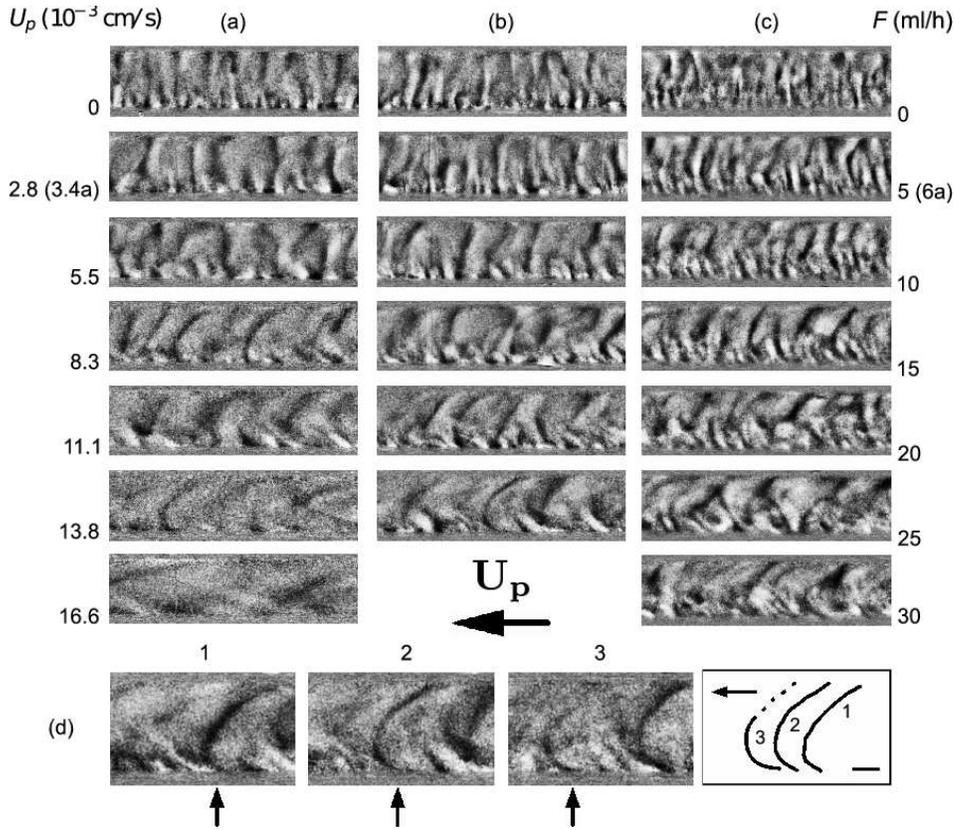}
\caption{Bioconvection patterns in tube B subject to imposed flow in the range $U_p=0$ to $16.6$ cm/s ($F=0$ to $35$ ml/h) for cell concentrations (a) $c=1.26\times106$ cells/ml, (b) $c=2.22\times106$ cells/ml and (c) $c=5.18\times106$ cells/ml. (As indicated, the second frame in (a) shows a frame for $3.4$ cm/s rather than $2.8$ cm/s). (d) For $U_p=13.8$ cm/s and concentration as in (b), we show three $5$ s interval snapshots of the dynamics of a bowed plume (indicated by arrows). The diagram to the right displays tracings of the plume (scale bar is $0.2$ cm).}\label{flowpatt}
\end{center}
\end{figure}

When flow was imposed using the syringe pump, patterns could be observed to be progressively distorted. Figure \ref{flowpatt} shows the patterns in tube B for (a) $c=1.26\times10^6$ cells/ml, (b) $c=2.22\times10^6$ cells/ml and (c) $c=5.18\times10^6$ cells/ml as the mean flow speed $U_p$ is increased in steps of $2.8$ cm/s from no flow to a maximum of $16.6$ cm/s (the flow is from right to left). The qualitative behaviour is generally similar for the different concentrations studied. In the absence of flow, the bioconvection plumes are parallel (on average) and slightly denser at the bottom. For small to moderate flow rates, most plumes still span the tube, but are seen to be curved so that the top and bottom of a plume are tilted at a characteristic angle (on average) to the vertical. The bowed plume shape is not symmetric for low concentrations, with plumes originating at the top stretching beyond the tube midpoint, before bending back at the bottom. Observation of the flowed pattern sequences reveals that plumes translate horizontally with the flow and that often high concentration pulses (blips) travel down a plume. As the flow rate is increased further, the pattern gets distorted more and more, with plumes tilting further away from the vertical at the top and bottom. The pattern appearance for higher flow rates is less orderly; many plumes break up, with less plumes spanning the tube. However, at high flow rates, the plumes becomes more symmetric: linear bottom standing plumes stretch to the tube midpoint. Remarkably, the plumes maintain a constant average angle at the top and bottom of the tube, which increases with flow rate, as plumes are tilted by the flow (see below). The reasons for this are nontrivial, see discussion. It should be remarked that the plume behaviour reported above is statistical: sequences show that individual plumes are not stable, meandering and either focusing or dispersing. For the highest flow rates all plumes fragment into very elongated structures making a large angle to the vertical and top and bottom plumes become fully symmetric about the tube midplane. As stated above, the phenomenology described is qualitatively similar for all concentrations studied, although there are quantitative differences.

\begin{figure}
\begin{center}
(a)\includegraphics[width=0.41\linewidth]{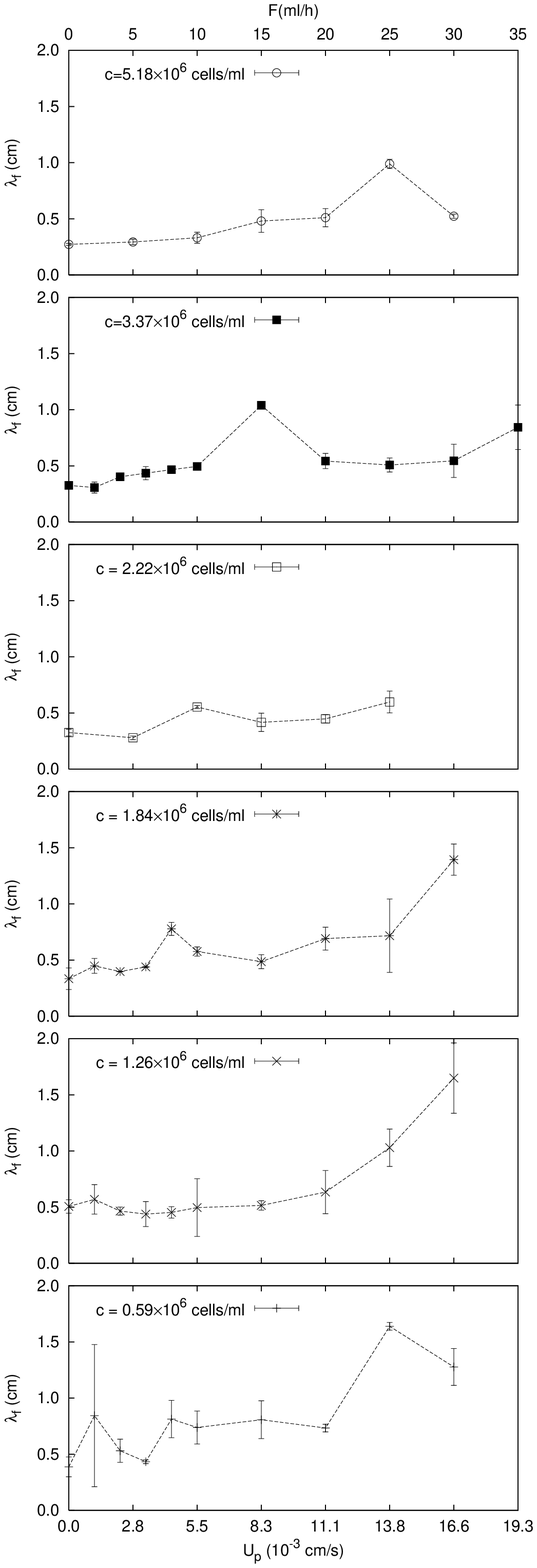}
(b)\includegraphics[width=0.41\linewidth]{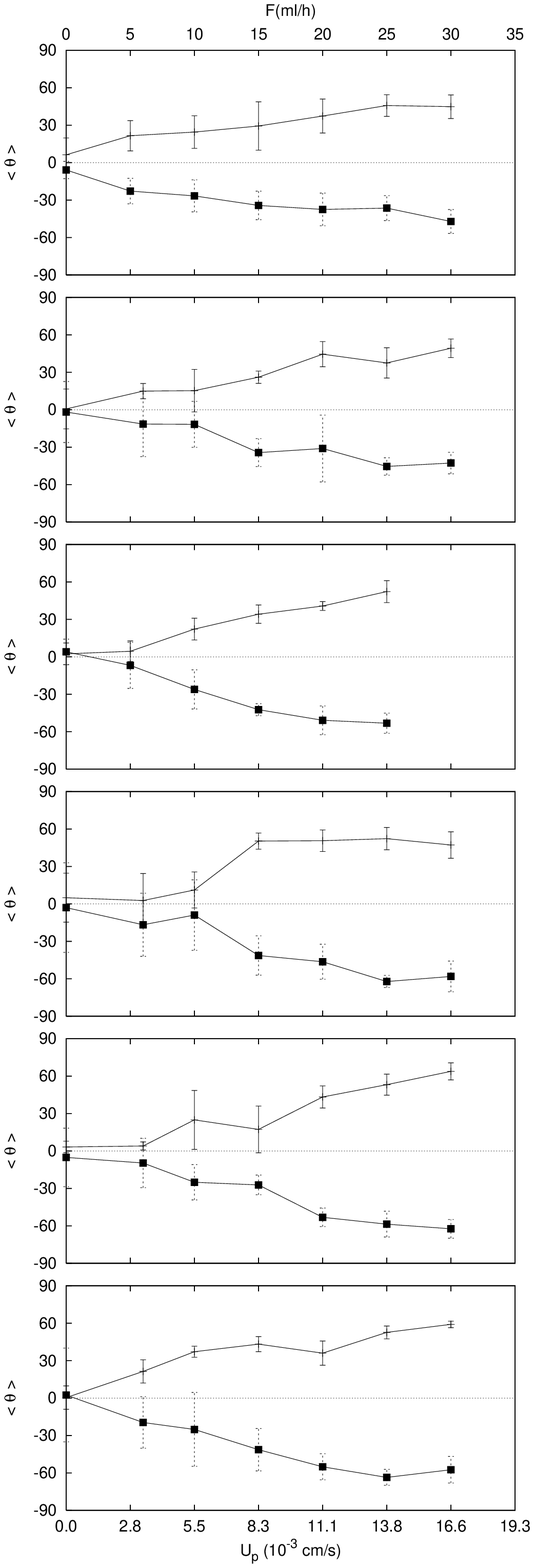}
\caption{(a) Final pattern wavelength and (b) average top (positive) and bottom (negative) plume angles to the vertical as a function of flow rate $F$ for concentrations in the range $c=0.59\times10^6-5.18\times10^6$ cells/ml, as indicated. The peaks in the wavelength around certain critical flow rates could be the signature of dynamic transitions in plume arrangement. The error bars shown are standard deviations.}\label{flowlam}
\end{center}
\end{figure}
\begin{figure}
\begin{center}
(a)\includegraphics[width=0.45\linewidth]{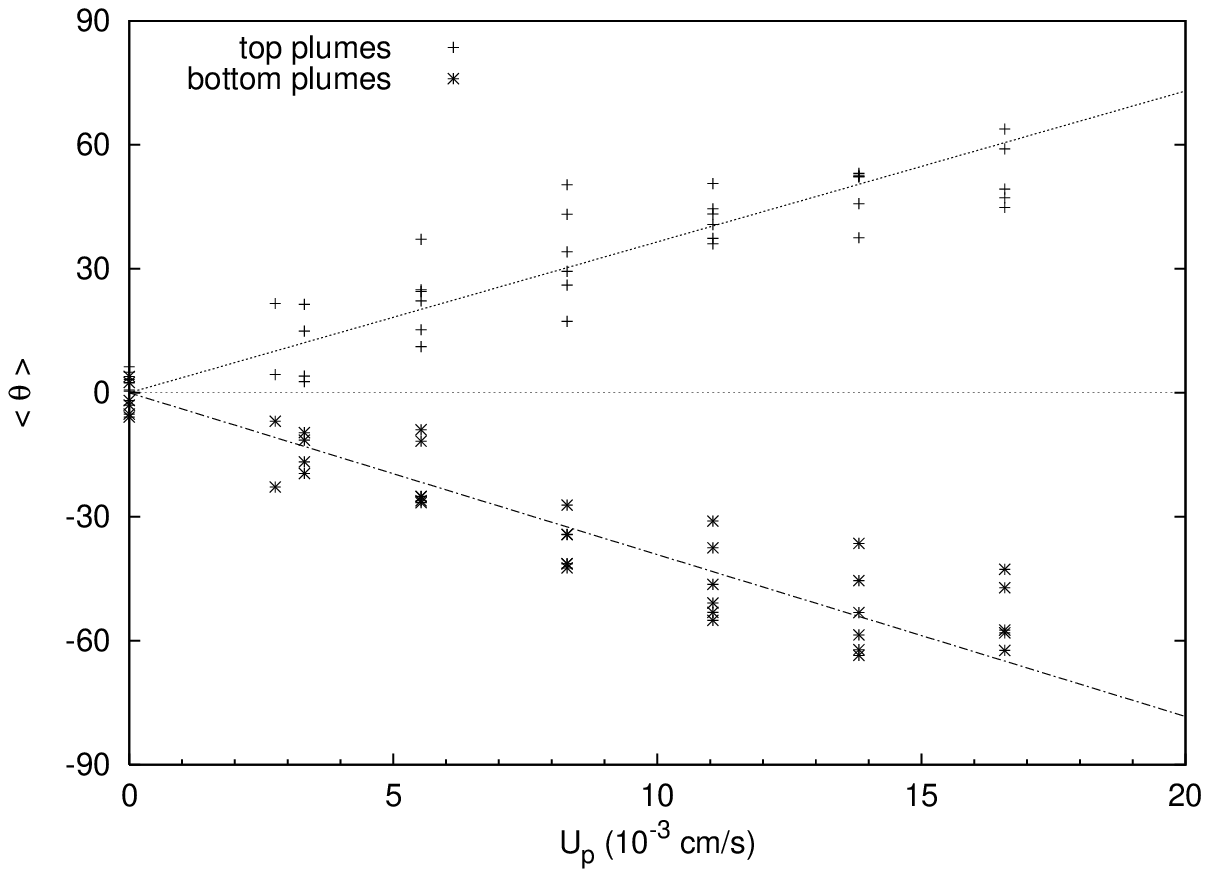}
(b)\includegraphics[width=0.45\linewidth]{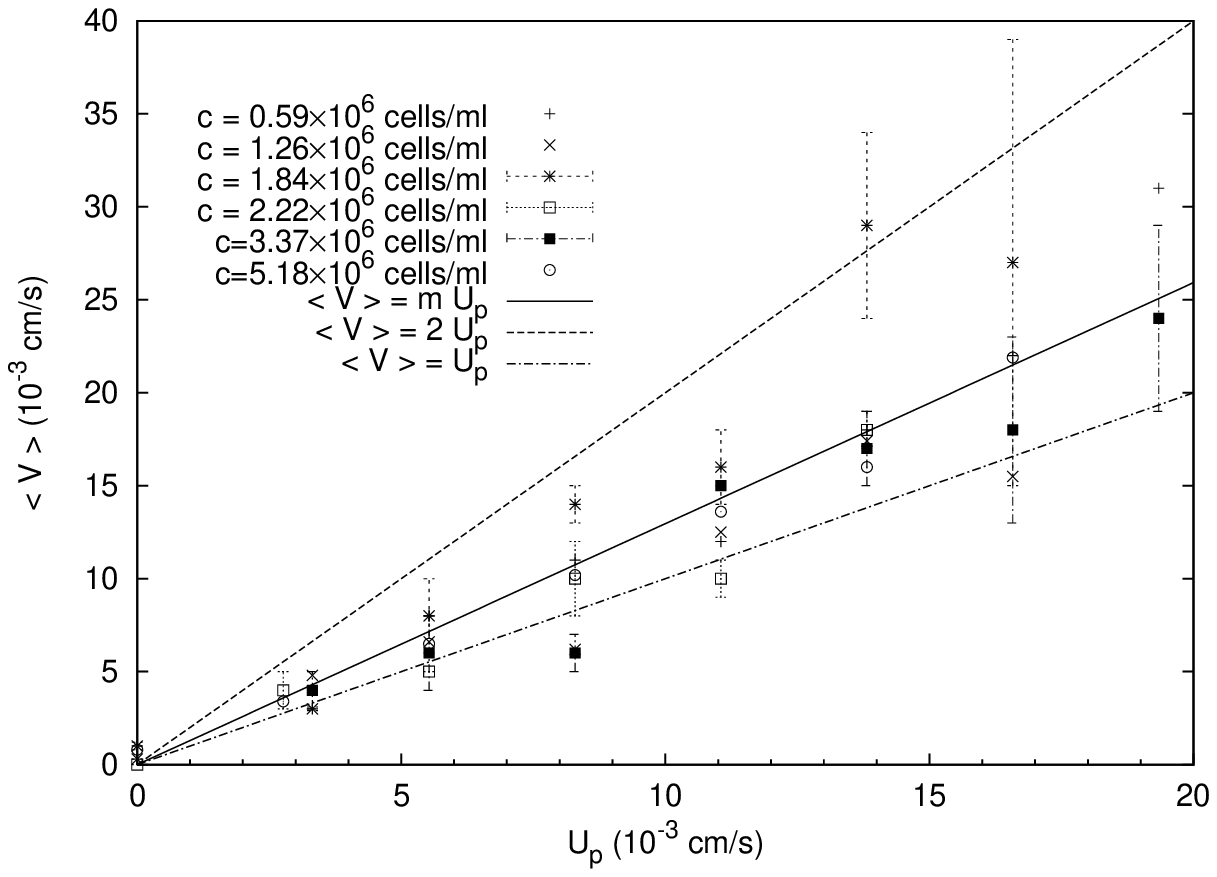}
\caption{Plot of (a) the average angles and (b) average plume drift speed as functions of the mean flow speed for all concentrations. Neither of these measures appears to depend strongly on concentration. Linear fits to data for all concentrations are also shown, and in the case of (b) the plume drift speed, the fit (solid line) is compared with the prediction for the mean (dot-dashed) and maximum (dashed) flow speeds for Poiseuille flow.}\label{flowangcollapse}
\end{center}
\end{figure}

To quantify the inclination of plumes, we measured the average angle, $\langle \theta \rangle$, a plume makes to the vertical when measured from the top and bottom of the tube (see methods). For top plumes, $\theta$ is defined as positive for tilt away from the vertical in the direction of the flow, while for bottom plumes it is negative. We see from figure \ref{flowlam}b, that plume tilt does not seem to vary sensitively on concentration. The apparent independence of $\langle \theta \rangle$ on concentration can be made clearer by plotting all concentrations on the same graph, as in figure \ref{flowangcollapse}a, where the change in $\langle \theta \rangle$ with flow rate $F$ has been fitted with a linear trend constrained to go through the origin (since we expect $\langle\theta\rangle=0$ for $F=0$ ml/h). The gradients obtained are very close in absolute value: $3.65\pm 0.11$ degrees s/cm  for the top plumes and $-3.91\pm0.13$  degrees s/cm for the bottom plumes. The average drift speed, $\langle V \rangle$, of the plumes was also measured. As shown in \ref{flowangcollapse}b, where all concentrations are plotted on the same graph, $\langle V \rangle$ appears to increase linearly with flow rate in a manner which appears generally independent of concentration. A constrained linear fit to the data, yields a value  $1.30 \pm 0.04$ for the change in average plume speed with mean flow speed. Also shown in \ref{flowangcollapse}b for comparison is the linear prediction for the maximum speed of the Poiseuille flow, $U_{poi}\equiv U=2 F/\pi a^2$ and for the equivalent plug flow $U_{p}=U/2$ (i.e.~the mean Poiseuille speed), giving gradients of 2 and 1, respectively. The data appears to fall between these two limits but closer to plug flow. We discuss the significance of this below.

A plot of the dominant pattern wavelength $\lambda$ as a function of flow rate $F$ (figure \ref{flowlam}a) shows a clear concentration dependence of the flowed pattern dynamics. For all the concentrations studied, as the flow rate is increased, $\lambda$ initially grows with flow rate. However, there are critical concentrations around which this growth is interrupted by a sudden rise and fall of $\lambda$, which quantitatively demarcates a dynamical transition in the pattern. These transitions represent a statistical change in the dynamics which is not immediately obvious from the frame sequences, however it is clear that these statistical changes in wavelength are associated with rearrangements of the plumes as the flow rate is increased.

\section{Discussion \label{disc}}

We have presented the first quantitative analysis of bioconvection patterns in horizontal tubes in the presence and absence of flow. A well mixed suspension of the green alga {\it C. augustae} becomes unstable to cell concentration fluctuations leading to a system of vertical plumes along the length of the tube.
With no flow imposed, the pattern sharpens with time, and the plume spacing decreases, consistent with observations by Wager (1911) of {\it Euglena viridis} and {\it Chlamydomonas} in similar tubes. To quantify these observations we Fourier analysed pattern sequences imaged from the side to measure initial and final dominant pattern wavelengths.  Both decrease slowly with concentration in the range $0.5\times10^6-4.5\times10^6$ cells/ml (except for a steeper drop for the initial wavelength for low concentrations in the smallest tube). Bees \& Hill (1997) observed a similar decrease with concentration for both the initial and final wavelength for bioconvection in shallow layers. They also found that the initial wavelength increased linearly with depth (see Czir\'{o}k \etal 2000), while the final wavelength appeared less sensitive to it.
The results here hint at a similar weak dependence on tube radius for a fixed concentration, but this cannot be inferred with confidence.  It would be interesting to perform experiments with a greater range of tube diameters to explore these trends.
It should be noted that we have studied projections onto a vertical plane of what are clearly three-dimensional patterns (see figure 1). The cross-sectional plume and flow structure may have a significant impact on the above interpretations of the results.

We have presented the first experimental study of bioconvection in the presence of imposed flows (with mean axial flow speeds that are comparable to bioconvective circulation).
We find that plumes are simply distorted for weak flows, but break as the flow rate increases. In this general phenomenology, the average plume drift speed $\langle V \rangle$ and inclination to the vertical $\langle \theta \rangle$, for a given flow rate, do not appear to depend appreciably on concentration in the range studied. However, plots of final pattern wavelength as a function of flow rate for different concentrations reveal that transitions in the pattern dynamics occur at critical flow rates that are sensitive to concentration. These transitions are more apparent from Fourier analysis than direct observation of the image sequences.

It will be useful to discuss our results in terms of approximations of the most recent model of bioconvection.  However, the solution of the full model is beyond the scope of this paper, so we shall resort to significant simplification later in this section to help explain our results.  We briefly summarize the model here, and refer the reader to Pedley \& Kessler (1992)
for a full exposition.  In a horizontal tube, the incompressible fluid flow velocity, $\bf{u}(\bf{r},t)$, is coupled to the cell density, $n(\bf{r},t)$, via the Navier-Stokes equations with a negative buoyancy term due to the cells, whilst a cell conservation equation describes flow dependent cell diffusion, swimming and advection (see Bees \& Croze 2010 for a vertical tube).  These flow dependent terms are obtained from a microscopic model of the hydrodynamics of swimming cells.

With no imposed flow, preliminary calculations (not shown) reveal the existence of a steady state profile in a horizontal tube, with most cells concentrated towards the top, in line with bioconvection between two horizontal boundaries.
Such a distribution may lead to an
overturning instability, as in analyses of other geometries (Hill \etal 1989; Bees \& Hill 1998).
Gyrotactic instabilities may also result where the cells are focused into plumes as a result of self-driven flow.
One may question whether
overturning or gyrotactic instabilities occur before the steady state has been approached.
When cells are placed in the tube, fluid motions decay with a time scale, $\tau_v=a^2/\nu$, where $a$  is the tube radius and $\nu$ is the kinematic viscosity. For the largest tube (tube A) $a\approx0.55$ cm, and taking $\nu=10^{-2}$ cm$^2$/s, we obtain $\tau_v=30$ s.  An unstable density profile will be established in the time it takes a cell to swim the lengthscale $a$, $\tau_o=a/V_s$, where $V_s$ is the cell swimming speed: with $V_s\approx10^{-2}$ cm/s, $\tau_o=55$ s. We can compare these estimates with the observed time for the onset of the initial instability, $\tau_i\approx30$ s ($c\approx 2.5\times10^6$; figure \ref{pattgrowthtime}). We see that $\tau_v\simeq\tau_i<\tau_o$; for the largest tube, the instability occurs roughly when mixing effects have decayed, but a little before the steady state profile has been established. A similar calculation yields $\tau_v\simeq \tau_i<\tau_o$ for tube B, and $\tau_v< \tau_i\simeq \tau_o$ for tube C (the steady state has been established before pattern onset with negligible mixing effects). Since $\tau_i$ increases for smaller $c$, patterns will likely be generated by an overturning instability
for dilute concentrations.

The observed evolution of patterns in the presence of flow results from coupling between the imposed suspension flux and bioconvective circulation.
The flow is laminar for the flow rates that we study; the Reynolds number, $\RRe=U a/\nu$, based on the maximum Poiseuille flow speed $U$, is $O(1)$.  The cells are advected horizontally, but plumes of falling cells redistribute fluid, modifying the flow.
In general, the secondary flows significantly alter the mean flow profile from its cell-less Poiseuille state, and the results suggest that the mean profiles are sometimes closer to plug flow.  For spherical cells, a simple balance of gravitational and viscous torques provides the angle to the vertical, $\psi$, at which gyrotactic cells swim in a shear flow with vorticity $\omega$, such that $\sin\psi=B\omega$, where $B=3.4$ s is the gyrotactic reorientation time (for an improved model see Pedley \& Kessler 1990, Bees \textit{et al.} 1998; stable orientation requires $\omega\le \omega_c =1/B=0.3$ s$^{-1}$, otherwise the cells tumble). The Poiseuille flow profile is $u(r)=U[1-(r/a)^2]$, giving vorticity $\omega(r)=(2 U/a^2)r$, whereas plug flow has $u(r)=U/2$ and $\omega(r)=0$. If Poiseuille flow is assumed with mean flow rates of $0.003$ to $0.02$ cm s$^{-1}$, the largest shear is smaller than $\omega_c$, with $\omega_{max}\equiv\omega(a)=0.03$ to $0.2$ s$^{-1}$ and thus maximum inclinations ranging from 6$^{\circ}$ to 43$^{\circ}$. Hence, we do not expect the formation of layers induced by gyrotactically trapped tumbling cells, as observed by Durham {\it et al.} (2009), except perhaps near the boundaries for flatter flow profiles.

The results indicate that for small to moderate flow rates, plumes remain whole but are distorted by the flow, starting at the top of the tube, tilting from the vertical and then bending back at the bottom of the tube.  This suggests that the circulation caused by descending plumes dominates that due to the imposed horizontal advection.
A striking feature of the plumes is that they appear to translate horizontally at a fixed average speed $\langle V \rangle$, mostly preserving their shape during their lifetime.
The progressive `breaking' of curved plume structures into angled linear plumes as the flow rate is increased further is harder to interpret.
Clearly stronger imposed flows disrupt bioconvective circulation to some degree.
One hypothesis is that for small flows the bioconvective circulation cells adjust and translate with the mean flow profile speed,
but when the flow becomes too strong the plumes stagger, splitting the large closed streamlines.
The dynamical transitions highlighted from measurements of the dominant pattern wavelength are intriguing.  From the results in figure \ref{flowlam} we infer that pattern transitions occur at higher flow rates for larger concentrations. This makes sense intuitively if we consider that for higher concentrations we expect a stronger cell-induced circulation, which will be more stable to perturbations by the imposed flow.
The resulting picture is that patterns are affected at all nonzero flow rates, but are severely disrupted, with dramatic transitions in the plume dynamics, only beyond a concentration dependent critical flow rate. The transitions only affect bioconvective circulation, and so plume arrangement, which would explain why the average angle $\langle \theta \rangle$ and average plume speed $\langle V \rangle$ do not depend on concentration.

\begin{figure}
\begin{center}
(a)\,\includegraphics[width=0.35\linewidth]{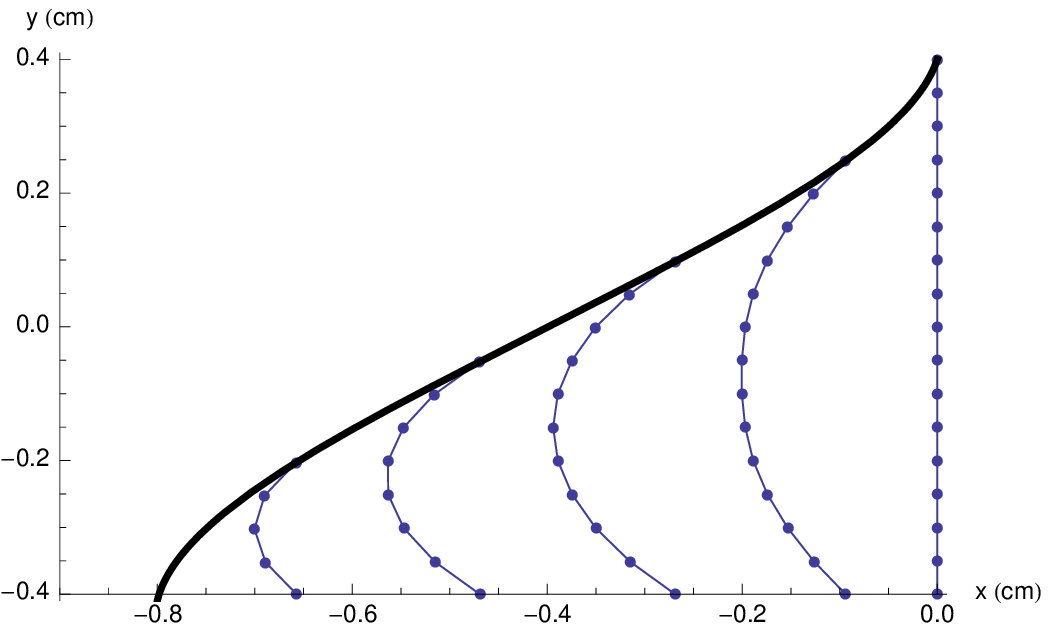}
\includegraphics[width=0.35\linewidth]{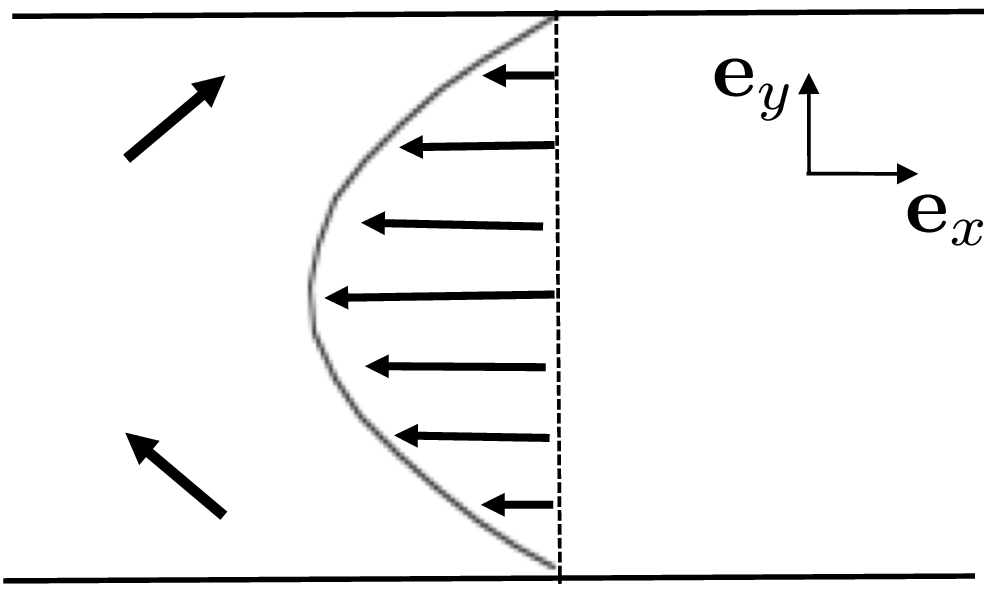}\\
(b)\,\includegraphics[width=0.35\linewidth]{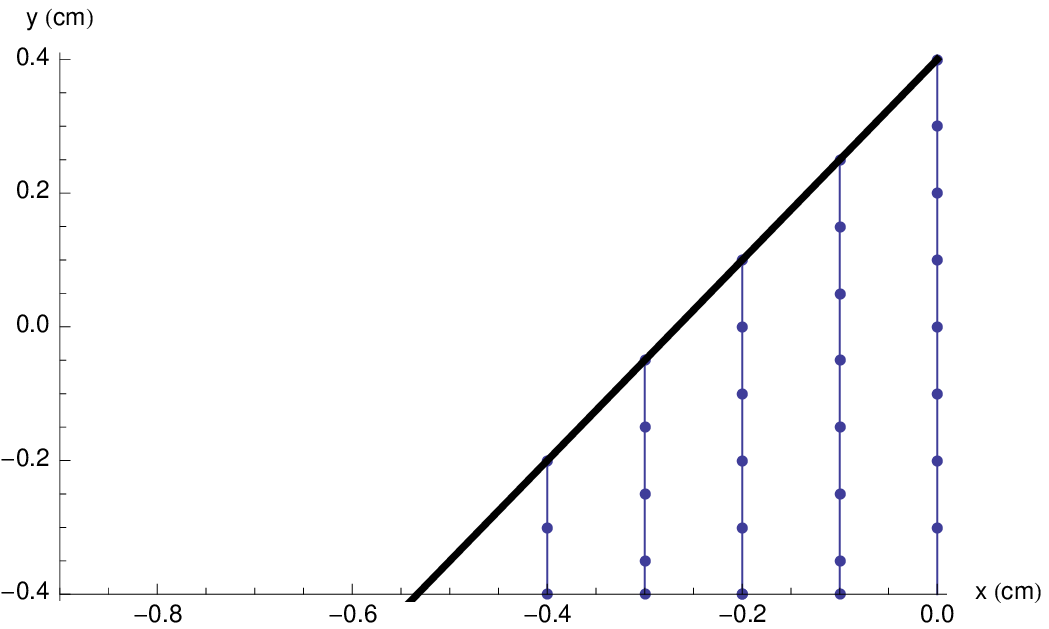}
\includegraphics[width=0.35\linewidth]{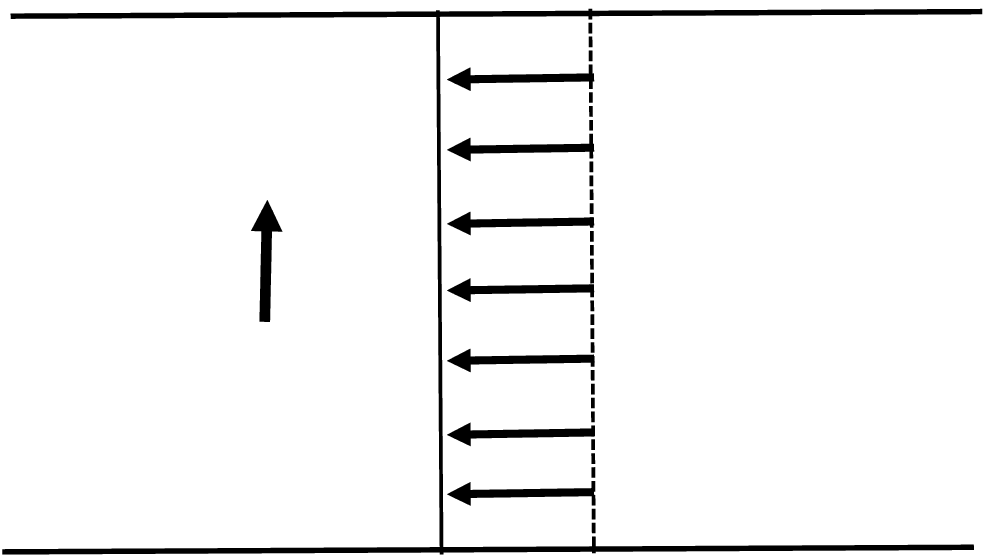}
\caption{Evolution of cell positions for (a) Poiseuille and (b) plug flows, according to the simplified description (see text). A line of cells (thin lines) released at equally spaced positions across the tube collapses and deforms while advected for (a), while it collapses vertically in (b). The stages in the evolution are $20$ s apart. The upper envelope (thick line) is the trajectory of cells released at the top of the tube, for the two flow scenarios.}\label{simpmodtraj}
\end{center}
\end{figure}

To help gain insight into cell dynamics in flows, we shall construct a much simplified description. Consider the plume structure in a vertical plane at the centre of the tube, with cartesian coordinates (x, y) as in figure \ref{simpmodtraj}.  We shall restrict attention to two-dimensional space for simplicity.
At its simplest, the position of a cell is subject to horizontal advection by the mean flow profile, ${u}(y)$, and vertical advection
in a plume, with constant speed $V$.  Furthermore, each cell swims at an angle to the vertical due to gyrotaxis with velocity
${\bf V}_c=V_s ( - B \omega(y) , \sqrt{1- (B \omega(y))^2})$, such that
\be\label{gyro}
\dot{x}(t)=u(y)-V_s B \omega(y) \mbox{~~~~and~~~~}
\dot{y}(t) =-V+ V_s \sqrt{1- B^2 \omega(y)^2}.
\ee
Figure \ref{simpmodtraj} presents the solution of (\ref{gyro}) for different initial conditions $(x_0, y_0)$, with $U=0.01$ cm/s, $V=0.0175$ cm/s, $V_s=0.01$ cm/s, $B=3.4$ s and $a=0.4$ cm. The thick lines indicate the path of a cell released at the top of the tube $(0, a)$, and is representative of a plume that is fed from the top. In Poiseuille flow (fig \ref{simpmodtraj}a) the path is curved due to the flow profile, and at the top and bottom shear induces diagonal gyrotactic reorientation (arrows). In plug flow (fig \ref{simpmodtraj}b) the cells on average move diagonally downward but swim upwards. Also shown (via a series of thin lines) is the evolution of a vertical line of cells, initially across the tube, at time intervals of $20$ s. For Poiseuille flow, the line translates with the horizontal component of the swimming velocity at the bottom as it collapses and deforms.
For plug flow the line remains vertical as it collapses and translates.
This simple description suggests that the behaviour of the plumes that we observe experimentally could result from a mean flow that is intermediate between Poiseuille and plug flows. In particular, the curved plumes are reminiscent of those for Poiseuille flow but the
curved plumes translate with a fixed shape as though they were in plug flow.  Furthermore, figure \ref{flowangcollapse}b demonstrates that plumes drift at a speed closer to plug flow than Poiseuille.
The model demonstrates that because advection by the flow and gravity are not collinear, gyrotactic reorientation by the imposed flow does not lead \textit{per se} to focusing (convergent trajectories) as in vertical tubes (Kessler 1985), although locally cells clearly drive the flow and will focus as a result.


The implications of the findings for the transport of cells in tubes are of particular interest. In vertical tubes, cells are subject to a modified Taylor-Aris dispersion: they are transported within a plume parallel to the flow, with advective, swimming and diffusive contributions (Bees \& Croze 2010). In horizontal tubes, the results presented here indicate that the situation is different, with cells transported in numerous plumes mostly translating at fixed speeds, $\langle V \rangle$.


To conclude, we have studied bioconvection in tubes with and without flow. The results demonstrate that even very weak laminar flows are sufficient to perturb bioconvection patterns and complex structures persist in the flow for larger flow rates, rather than flow simply having a mixing effect. The presence of the cells appears to modify the mean flow to be more like plug flow than Poiseuille.
Future experimental investigations should measure the (mean) horizontal flows using particle tracking or PIV.
Tubular bioreactor designs comprise arrangements of horizontal, vertical and/or inclined tubes in which cells are can be in transient turbulent or turbulent flow to maximize mixing and equalize light exposure (Grima {\it et al.} 2001, 2003;  Garc\'{i}a-Gonz\'{a}lez \etal 2005; Chisti 2007). It is important to establish the most efficient flow rates for maximum biomass production with minimal energy consumption, and how the optical properties of bioconvecting suspensions (transmittance) change with the flow and couple to growth.
Towards this end, insights into the transition to turbulence in the presence of cells will be invaluable, as will an understanding of how cells are arranged within a developed turbulent pipe flow (see Lewis 2003).
The experiments presented here represent a first step in this direction.
Theoretically, bioconvection in horizontal tubes has not yet been tackled; the main challenge is to predict the average inclination and speed of plumes, as well as the flow transitions we have observed, as functions of concentration and tube diameter. An open question is how to predict the effective transport properties of swimming cells in laminar and turbulent regimes in tubes of arbitrary orientation.

\ack

MAB and OAC acknowledge support from the EPSRC (EP/D073398/1). EEA acknowledges support from the Pakistan Air Force and the Government of Pakistan.

\section*{References}

\begin{harvard}

\item
Bearon, R.~N. \& Gr\"{u}nbaum, D. 2006 Bioconvection in a stratified
  environment: Experiments and theory.
\textit{Phys. Fluids}, \textbf{18}, 127102.

\item
Bees, M.~A. \& Croze, O.~A. 2010 Dispersion of biased micro--organisms in a
  fluid flowing through a tube.
\textit{Proc. R. Soc. A}, \textbf{466}, 1067--1070.

\item
Bees, M.~A. \& Hill, N.~A. 1997 Wavelengths of bioconvection patterns.
\textit{J. Exp. Biol.}, \textbf{200}, 1515--1526.

\item
Bees, M.~A. \& Hill, N.~A. 1998 Linear bioconvection in a suspension of
  randomly swimming, gyrotactic micro-organisms.
\textit{Phys. Fluids}, \textbf{10}, 1864--1881.

\item
Bees, M.~A. \& Hill, N.~A. 1999 Non-linear bioconvection in a deep suspension
  of gyrotactic micro-organisms.
\textit{J. Math. Biol.}, \textbf{38}, 135--168.

\item
Bees, M.~A.,  Hill, N.~A. \& Pedley, T.~J. 1998 Analytical approximations for the orientation distribution of small dipolar particles in steady shear flows.
\textit{J. Math. Biol.}, \textbf{36}, 269--298.

\item
Berg, H.~C. 2004 \emph{{\it E. coli} in motion}.
\newblock Springer.

\item
Chandrasekar, S. 1961 \emph{Hydrodynamic and hydromagnetic stability}.
\newblock Clarendon-Press (Oxford).

\item
Childress, W. S., Levandowsky, M. \& Spiegel, E.~A. 1975 Pattern formation in a
  suspension of swimming micro-organisms.
\textit{J. Fluid Mech.}, \textbf{69}, 595--613.

\item
Chisti, Y. 2007 Biodiesel from microalgae.
\textit{Biotechnol. Adv.}, \textbf{25}, 294--306.

\item
Czir\'{o}k, A., J\'{a}nosi, I.~M. \& Kessler, J.~O. 2000 Bioconvective dynamics: dependence on organism behaviour.
\textit{J. Exp. Biol.}, \textbf{203}, 3345--3354.

\item
Durham, W.~M., Kessler, J.~O. \& Stocker, R. 2009 Disruption of vertical
  motility by shear triggers formation of thin phytoplankton layers.
\textit{Science}, \textbf{323}, 1067--1070.

\item
Foster, K.~W. \& Smyth, R.~D. 1980 Light antennas in phototactic algae.
\textit{Microbiol. Rev.}, \textbf{44}, 572--630.

\item
Grima, E.~M., Belarbi, E.-H., Fern\'{a}ndez, F.~G.~A., Medina, A.~R. \& Chisti,
  Y. 2003 Recovery of microalgal biomass and metabolites: process options and
  economics.
\textit{Biotechnol. Adv.}, \textbf{20}, 491.

\item
Grima, E.~M., Fern\'{a}ndez, J., Fern\'{a}ndez, F.~G.~A. \& Chisti, Y. 2001
  Tubular photobioreactor design for algal cultures.
\textit{J. Biotechnol.}, \textbf{92}, 113--131.

\item
Gr\"{u}nbaum, D. 2009 Peter principle packs a peck of phytoplankton.
\textit{Science}, \textbf{323}(5917), 1022--1023.

\item
Hill, N.~A. \& Pedley, T.~J. 2005 Bioconvection.
\textit{Fluid Dyn. Res.}, \textbf{37}, 1--20.

\item
Hill, N.~A., Pedley, T.~J. \& Kessler, J.~O. 1989 The growth of bioconvection
  patterns in a uniform suspension of gyrotactic micro-organisms in a layer of
  finite depth.
\textit{J. Fluid Mech.}, \textbf{208}, 509--543.

\item
J\'{a}nosi, I.~M., Kessler, J.~O. \& Horv\'{a}th, V.~K. 1998 Onset of
  bioconvection in suspensions of {\it bacillus subtilis}.
\textit{Phys. Rev. E}, \textbf{58}, 4793--4800.

\item
Kessler, J.~O. 1985 Hydrodynamic focusing of motile algal cells.
\textit{Nature}, \textbf{313}, 218.

\item
Kessler, J.~O., Hill, N.~A. \& H\"ader, D.-P. 1992 Orientation of swimming
  flagellates by simulatenously acting external factors.
\textit{J. Phycol.}, \textbf{28}, 816--822.

\item
Levandowsky, M., Childress, S., Spiegel, E.~A. \& Hunter, S.~H. 1975 A mathematical model for pattern formation by swimming micro-organisms.
\textit{J. Protozool.}, \textbf{22}, 296--306.

\item
Lewis, D.~M. 2003 The orientation of gyrotactic spheroidal micro-organisms in a
  homogeneous isotropic turbulent flow.
\textit{Proc. R. Soc. A}, \textbf{459}, 1293--1323.

\item
Garc\'{i}a-Gonz\'{a}lez, M., Moreno, J., Manzano, J.~C., Florencio, F.~J. \& Guerrero, M.~G. 2005
  Production of {\it dunaliella salina} biomass rich in 9-{\it
  cis}-$\beta$-carotene and lutein in a closed tubular bioreactor.
\textit{J. Biotechnol.}, \textbf{115}, 81--90.

\item
Melis, A. \& Happe, T. 2001 Hydrogen production: green algae as a source of
  energy.
\textit{Plant Physiol.}, \textbf{127}, 740--748.

\item
Nultsch, W., Throm, G. \& von Rimscha, I. 1971 Phototaktische untersuchungen an
  {\it Chlamydomonas reinhardii} dangeard in homokontinuierlicher kultur.
\textit{Arch. Mikrobiol.}, \textbf{80}, 351--369.

\item
Pedley, T.~J. \& Kessler, J.~O. 1990 A new continuum model for suspensions of
  gyrotactic micro-organisms.
\textit{J. Fluid Mech.}, \textbf{212}, 155--182.

\item
Pedley, T.~J. \& Kessler, J.~O. 1992 Hydrodynamic phenomena in suspensions of
  swimming micro-organisms.
\textit{Annu. Rev. Fluid Mech.}, \textbf{24}, 313--358.


\item
Pr\"{o}schold, T., Marin, B., Schl\"{o}sser, U.~G. \& Melkonian, M. 2001
  Molecular phylogeny and taxonomic revision of {\it Chlamydomonas} (Chlorophyta). I.
  Emendation of {\it Chlamydomonas} Ehrenberg  and {\it Chloromonas} Gobi, and description
  of {\it Oogamochlamys} gen. nov. and {\it Lobochlamys} gen. nov.
\textit{Protist}, \textbf{152}, 265--300.

\item
Schl\"{o}sser, U.~G. 1997 Additions to the culture collection of algae since
  1994.
\textit{Bot. Acta}, \textbf{110}, 424--429.

\item
Wager, H. 1911 On the effect of gravity upon the movements and aggregation of
  {\it euglena viridis}, ehrb., and other micro--organisms.
\textit{Phil. Trans. R. Soc. Lond. B}, \textbf{201}, 333--390.

\item
Weibel, D.~B., Garstecki, P., Ryan, D., DiLuzio, W.~R., Mayer, M., Seto, J.~E.
  \& Whitesides, G.~M. 2005 Microoxen: Microorganisms to move microscale loads.
\textit{Proc. Natl. Acad. Sci USA}, \textbf{102}, 11963.

\item
Willis, A.~P., Peixinho, J., Kerswell, R.~R. \& Mullin, T. 2008 Experimental
  and theoretical progress in pipe flow transition.
\textit{Phil. Trans. R. Soc. Lond. A}, \textbf{366}, 2671--2684.

\end{harvard}

\end{document}